\documentclass[12pt]{article}

\usepackage{graphicx}
\usepackage{amsmath}
\usepackage{amssymb}
\numberwithin{equation}{section}

\newcommand{\bs}[1]{\boldsymbol{#1}}
\newcommand{\Y}{\!\not\hspace{-0.5mm}Y}
\newcommand{\calY}{\!\not\hspace{-1mm}{\cal Y}}
\begin{document}

\begin{titlepage}
\title{
\hfill\parbox{4cm}
{\normalsize KEK-TH-1102\\MIT-CTP-3784\\
{\tt hep-th/0611113}}\\
\vspace{1cm}
{\bf On LCSFT/MST Correspondence}
}
\author{
{\sc Isao Kishimoto}{}\thanks
{{\tt ikishimo@post.kek.jp}}\\[6pt]
{\it High Energy Accelerator Research Organization (KEK),}\\
{\it Tsukuba 305-0801, Japan}\\[12pt]
{\sc Sanefumi Moriyama}{}\thanks
{{\tt moriyama@math.nagoya-u.ac.jp}}\\[6pt]
{\it Graduate School of Mathematics, Nagoya University,}\\
{\it Nagoya 464-8602, Japan}\,{}\thanks{Permanent address}\\[6pt]
{\it Center for Theoretical Physics, 
Massachusetts Institute of Technology,}\\
{\it Cambridge, MA02139, USA}\\
}
\date{\normalsize November, 2006}
\maketitle
\thispagestyle{empty}

\begin{abstract}
\normalsize

It was suggested that light-cone superstring field theory (LCSFT) and
matrix string theory (MST) are closely related.
Especially the bosonic twist fields and the fermionic spin fields in MST
correspond to the string interaction vertices in LCSFT.
Since CFT operators are characterized by their OPEs, in our previous
work we realized the most important OPE of the twist fields by computing
contractions of the interaction vertices using the bosonic cousin of
LCSFT.
Here using the full LCSFT we generalize our previous work into the
realization of OPEs for a vast class of operators.

\end{abstract}

\end{titlepage}

\section{Introduction}
Search for a fundamental field-theoretical formulation of string theory
began in the old days of dual resonance models.
However, none of these attempts have been completed yet for the
superstring theory.

The light-cone superstring field theory (LCSFT) is the most successful
formulation thus far and was first constructed in \cite{GS,GSB}.
The starting point is the Green-Schwarz action in the light-cone
gauge.
Out of the action we can construct the free Hamiltonian $H_0$ and two
free supercharges $Q_0^{\dot a}, \tilde Q_0^{\dot a}$, which satisfy the
supersymmetry algebra:
\begin{align}
\{Q_0^{\dot{a}},Q_0^{\dot{b}}\}
=\{\tilde Q_0^{\dot{a}},\tilde Q_0^{\dot{b}}\}
&=2\delta^{\dot{a}\dot{b}}H_0,
\label{susyalg1}\\
[Q_0^{\dot{a}},H_0]=[\tilde Q_0^{\dot{a}},H_0]&=0.
\label{susyalg2}
\end{align}
The interaction terms are added to these charges so that the total
charges satisfy the same supersymmetry algebra order by order.
Namely, we construct the interaction terms by replacing the free charges
$H_0, Q_0^{\dot a}, \tilde Q_0^{\dot a}$ in \eqref{susyalg1},
\eqref{susyalg2} by the full order charges
\begin{align}
H&=H_0+g_sH_1+\cdots,\\
Q^{\dot a}&=Q^{\dot a}_0+g_sQ^{\dot a}_1+\cdots,\\
\tilde{Q}^{\dot a}&=\tilde{Q}^{\dot a}_0+g_s\tilde{Q}^{\dot a}_1+\cdots,
\end{align}
and determining the interaction terms order by order.
The result for the first order interaction terms is
\begin{align}
|H_1\rangle_{123}
&=Z^i\bar Z^jv^{ji}(Y)|V\rangle_{123},
\label{SFTH1}\\
|Q_1^{\dot{a}}\rangle_{123}
&=\bar Z^is^{i\dot{a}}(Y)|V\rangle_{123},\\
|\tilde Q_1^{\dot{a}}\rangle_{123}
&=Z^i\tilde s^{i\dot{a}}(Y)|V\rangle_{123}.
\end{align}
Here $|V\rangle_{123}$ is the kinematical three-string interaction
vertex constructed by the overlapping condition and $Z^i$ $(\bar Z^i)$
is the holomorphic (anti-holomorphic) part of the bosonic momentum at
the interaction point, whose divergence is regularized:
\begin{align}
\Bigl(P^{(1)i}+\frac{1}{2\pi\alpha_1}X^{(1)i\prime}\Bigr)(\sigma_1)
|V\rangle_{123}
\sim\frac{1+ i\epsilon(\sigma_{\rm int}-\sigma_1)}
{4\pi\sqrt{-\alpha_{123}}
\sqrt{|\sigma_{\rm int}-\sigma_1|}}Z^i|V\rangle_{123},
\label{Z}
\end{align}
with $\alpha_r=p^+_{r}$, $\alpha_{123}=\alpha_1\alpha_2\alpha_3$ and
$\sigma_{\rm int}$ being the interaction point.
We take the range of $\sigma_1$ to be
$-\pi\alpha_1\le\sigma_1\le\pi\alpha_1$, which implies
$\sigma_{\rm int}=\pi\alpha_1$ for $\sigma_1\sim\pi\alpha_1$ and
$\sigma_{\rm int}=-\pi\alpha_1$ for $\sigma_1\sim-\pi\alpha_1$,
respectively.
$\epsilon(x)$ is the step function and defined by $\epsilon(x)=+1$
($-1$) for $x>0$ ($x<0$).
Similarly, $Y^a$ is the regularization of the fermionic momentum at the
interaction point:
\begin{align}
\lambda^{(1)a}(\sigma_1)|V\rangle_{123}
\sim\frac{1}{4\pi\sqrt{-\alpha_{123}}
\sqrt{|\sigma_{\rm int}-\sigma_1|}}Y^a
|V\rangle_{123},
\label{Y}
\end{align}
and the prefactors $v^{ji}(Y)$, $s^{i\dot{a}}(Y)$ and
$\tilde{s}^{i\dot{a}}(Y)$ are functions of $Y^a$ and were originally
given in complicated forms.
Here we find that these functions can be simply put into the hyperbolic
functions:
\begin{align}
v^{ji}(Y)&=\bigl[\cosh\Y\bigr]^{ij}
=\delta^{ij}+\frac{1}{2!}(\Y^2)^{ij}
+\frac{1}{4!}(\Y^4)^{ij}+\frac{1}{6!}(\Y^6)^{ij}+\frac{1}{8!}(\Y^8)^{ij},
\label{v}\\
s^{i\dot a}(Y)&=\sqrt{-\alpha_{123}}\bigl[\sinh\Y\bigr]^{\dot{a}i}
=\sqrt{-\alpha_{123}}\Bigl[(\Y)^{\dot{a}i}
+\frac{1}{3!}(\Y^3)^{\dot{a}i}+\frac{1}{5!}(\Y^5)^{\dot{a}i}
+\frac{1}{7!}(\Y^7)^{\dot{a}i}\Bigr],
\label{s}\\
\tilde{s}^{i\dot a}(Y)
&=i\sqrt{-\alpha_{123}}\bigl[\sinh\Y\bigr]^{i\dot{a}}
=i\sqrt{-\alpha_{123}}\Bigl[(\Y)^{i\dot{a}}
+\frac{1}{3!}(\Y^3)^{i\dot{a}}+\frac{1}{5!}(\Y^5)^{i\dot{a}}
+\frac{1}{7!}(\Y^7)^{i\dot{a}}\Bigr],
\label{tildes}
\end{align}
where we have constructed gamma matrices with spinor indices
$\hat\gamma^a$ out of the standard gamma matrices with the vector
indices $\gamma^i$ using the triality of $SO(8)$ and defined $\Y$ as
\begin{align}
\Y=\sqrt{\frac{2}{-\alpha_{123}}}\eta^*Y^a\hat{\gamma}{}^a
=\begin{pmatrix}0&\Y_{i\dot a}\\
\Y_{\dot{a}i}&0\end{pmatrix},
\label{Yslash}
\end{align}
with $\eta^*=e^{-\frac{i\pi}{4}}$.
Note that the indices of the functions in \eqref{v}--\eqref{tildes} are
consistent because cosh is an even function while sinh is an odd one.
For more details of the prefactors, see Appendix A.
As pointed out in \cite{GS}, these quantities satisfy the following
Fourier identities:
\begin{align}
\bigl[\cosh\Y\bigr]^{ij}
&=\left(\alpha_{123}\over 2\right)^4\int d^8\phi\,
e^{{2\over \alpha_{123}}\phi^aY^a}
\bigl[\cosh\!\not\hspace{-1mm}\phi\bigr]^{ji},
\label{fourier1}\\
\bigl[\sinh\Y\bigr]^{\dot{a}i}
&=-i\left(\alpha_{123}\over 2\right)^4\int d^8\phi\,
e^{{2\over \alpha_{123}}\phi^aY^a}
\bigl[\sinh\!\not\hspace{-1mm}\phi\bigr]^{i\dot{a}},\\
\bigl[\sinh\Y\bigr]^{i\dot{a}}
&=i\left(\alpha_{123}\over 2\right)^4\int d^8\phi\,
e^{{2\over \alpha_{123}}\phi^aY^a}
\bigl[\sinh\!\not\hspace{-1mm}\phi\bigr]^{\dot{a}i},
\label{fourier3}
\end{align}
which will play important roles in our following computation.
The program of constructing the interaction terms is successful at the
first order, though it is too complicated to proceed to higher orders.

Recently, there is a significant breakthrough in the construction
\cite{DM}.
The point is to relate LCSFT to another formulation of the superstring
theory known as matrix string theory (MST) \cite{MSSBS,DVV}.
MST stems from the matrix formulation of light-cone quantization of
M-theory \cite{BFSS} and takes the form of the maximally supersymmetric
Yang-Mills theory.
To relate MST to the perturbative string, we note that the Yang-Mills
coupling constant $g_{\rm YM}$ is related to the string coupling
constant $g_{\rm s}$ and the string length $\sqrt{\alpha'}$ by
$g_{\rm YM}^{-1}\sim g_{\rm s}\sqrt{\alpha'}$.
Hence, the free string limit corresponds to the IR limit and the first
order interaction term to the least irrelevant operator.
{}From the requirement of the dimension counting and the locality of the
interaction, we expect that the first order interaction term is a
dimension three operator constructed essentially out of the bosonic
twist fields and the fermionic spin fields.
The interaction term of MST is proposed to be \cite{DVV}
\begin{align}
H_1=\sum_{m,n}\int d\sigma
\bigl([\tau^i\bar{\tau}^j]\Sigma^i\bar{\Sigma}^j\bigr)_{m,n},
\label{MSTH1}
\end{align}
where $\tau^i(z)$ is the excited ${\mathbb Z}_2$ twist field defined as
\begin{align}
\partial X^i(z)\cdot\sigma(0)
\sim\frac{1}{\sqrt{z}}\tau^i(0),
\label{tau}
\end{align}
with $\sigma(z)$ being the elementary ${\mathbb Z}_2$ twist field.
Since the holomorphic part of the twist field does not completely
decouple from the anti-holomorphic part due to the zero mode
contribution, we combine them with the square parenthesis in
\eqref{MSTH1}.
$\Sigma^i(z)$ and $\Sigma^{\dot{a}}(z)$ (which will appear later) are
the spin fields for the Green-Schwarz fermion $\theta^a(z)$.
The indices $m$ and $n$ of the twist fields denote the string bits where
the ``exchange'' interaction takes place.

Now we can find a close analogy between \eqref{SFTH1} and \eqref{MSTH1}
and between \eqref{Z} and \eqref{tau}, if we regard
$[\sigma\bar\sigma](z,\bar{z})$ as $|V\rangle_{123}$ and
$[\tau^i\bar\tau^j](z,\bar{z})$ as $Z^i\bar{Z}^j|V\rangle_{123}$. 
Following this analogy between LCSFT and MST, two supercharges of MST
were written down explicitly in \cite{M}:
\begin{align}
Q_1^{\dot{a}}=\sum_{m,n}\int d\sigma
\bigl([\sigma\bar{\tau}^i]\Sigma^{\dot{a}}\bar{\Sigma}^i\bigr)_{m,n},
\quad
\tilde{Q}_1^{\dot{a}}=\sum_{m,n}\int d\sigma
\bigl([\tau^i\bar{\sigma}]\Sigma^i\bar{\Sigma}^{\dot{a}}\bigr)_{m,n},
\end{align}
and the supersymmetry algebra was checked.
These arguments of supercharges are consistent with the pioneering
argument in \cite{DVV} and with the relation between LCSFT and MST
proposed in \cite{DM}.

After observing the analogy between LCSFT and MST, we would like to
establish the correspondence next.
Since operators in conformal field theory are characterized by their
OPEs, we need to reproduce the OPEs in terms of LCSFT in order to
completely confirm the correspondence.
In our previous work \cite{KMT}, we assumed the correspondence between
the bosonic cousin of LCSFT
\begin{align}
|H_1^{\rm boson}\rangle_{123}=|V^{\rm b}\rangle_{123},
\label{bLCSFT}
\end{align}
and the ``bosonic'' version of MST \cite{Rey}
\begin{align}
H_1^{\rm boson}
=\sum_{m,n}\int d\sigma\bigl([\sigma\bar\sigma]\bigr)_{m,n},
\label{bMST}
\end{align} 
and reproduced the OPE of the twist field
\begin{align}
[\sigma\bar\sigma](z,\bar z)\cdot[\sigma\bar\sigma](0,0)
\sim\frac{1}{\bigl[|z|^{1/4}(\log|z|)^{1/2}\bigr]^{d-2}},
\label{twosigma}
\end{align}
by computing the corresponding contractions of the string interaction
vertices in the bosonic LCSFT.
On the LCSFT side, it is well-known that the first order interaction
vertex in the bosonic LCSFT \eqref{bLCSFT} takes the same form as that
in the full LCSFT, if we drop the prefactors and pick up the bosonic
sector in the kinematical interaction vertex $|V\rangle_{123}$.
The superscript ``b'' in \eqref{bLCSFT} denotes that we pick up the
bosonic sector.
On the MST side, of course there is no nonperturbative justification for
the bosonic MST, but the interaction term \eqref{bMST} looks natural
considering the correct dimension of the operators $[\sigma\bar\sigma]$
and the correspondence between $[\sigma\bar\sigma](z,\bar{z})$ and
$|V\rangle_{123}$.

Since the string interaction vertex describes both the processes of
one string splitting into two strings and two strings joining into one
string, there are two diagrams which correspond to the OPE
\eqref{twosigma}.
One of them is a tree diagram with two incoming strings and two
outgoing strings of the same string lengths, while the other is a
one-loop diagram with one incoming string and one outgoing string.
(See Fig.~1 and Fig.~2 in the following sections.)
For the two realizations of the OPE, the corresponding contractions of
the interaction vertices are given by
\begin{align}
&{}_{36}\langle R^{\rm b}|
e^{-\frac{T}{|\alpha_3|}(L_0^{(3)}+\bar{L}_0^{(3)})}
|V^{\rm b}\rangle_{123}|V^{\rm b}\rangle_{456},\nonumber\\
&{}_{14}\langle R^{\rm b}|{}_{25}\langle R^{\rm b}|
e^{-\frac{T}{\alpha_1}(L_0^{(1)}+\bar{L}_0^{(1)})
-\frac{T}{\alpha_2}(L_0^{(2)}+\bar{L}_0^{(2)})}
|V^{\rm b}\rangle_{123}|V^{\rm b}\rangle_{456}.
\label{RVV}
\end{align}
We found that both the contractions of the string vertices give exactly
the same singular behavior expected from the OPE \eqref{twosigma}.

Though the correspondence seems to work well also in the bosonic case,
the action \eqref{bMST} still lacks justification.
In this paper we would like to return to our original motivation of
investigating the correspondence in the supersymmetric case, where we
typically have to deal with the prefactors.
First, we repeat the computation of the bosonic sector.
In order to reproduce the OPE
\begin{align}
[\tau^i\bar\sigma](z,\bar{z})\cdot[\tau^k\bar\sigma](0,0)
\sim\frac{\delta^{ik}}{z^2\bar{z}(\log|z|)^4},
\end{align}
in terms of LCSFT, from the correspondence dictionary we need to
consider two contractions:
\begin{align}
&{}_{36}\langle R^{\rm b}|
e^{-\frac{T}{|\alpha_3|}(L_0^{(3)}+\bar{L}_0^{(3)})}
Z_{123}^i|V^{\rm b}\rangle_{123}
Z_{456}^k|V^{\rm b}\rangle_{456},\nonumber\\
&{}_{14}\langle R^{\rm b}|{}_{25}\langle R^{\rm b}|
e^{-\frac{T}{\alpha_1}(L_0^{(1)}+\bar{L}_0^{(1)})
-\frac{T}{\alpha_2}(L_0^{(2)}+\bar{L}_0^{(2)})}
Z_{123}^i|V^{\rm b}\rangle_{123}
Z_{456}^k|V^{\rm b}\rangle_{456},
\label{RVVb}
\end{align}
for the tree diagram and the one-loop diagram.
After the computation in the bosonic sector, we shall proceed to the
fermionic sector.
Among various OPEs of the spin fields, we would like to concentrate on
the realization of the OPE
\begin{align}
\Sigma^i(z)\bar\Sigma^j(\bar{z})\cdot\Sigma^k(0)\bar\Sigma^l(0)
\sim\frac{\delta^{ik}\delta^{jl}}{|z|^2}.
\end{align}
This OPE corresponds to the fermionic parts of the contractions between
two first order Hamiltonians $|H_1\rangle$ \eqref{SFTH1} in LCSFT:
\begin{align}
&{}_{36}\langle R^{\rm f}|
e^{-\frac{T}{|\alpha_3|}(L_0^{(3)}+\bar{L}_0^{(3)})}
\bigl[\cosh\Y_{123}\bigr]^{ij}|V^{\rm f}\rangle_{123}
\bigl[\cosh\Y_{456}\bigr]^{kl}|V^{\rm f}\rangle_{456},\nonumber\\
&{}_{14}\langle R^{\rm f}|{}_{25}\langle R^{\rm f}|
e^{-\frac{T}{\alpha_1}(L_0^{(1)}+\bar{L}_0^{(1)})
-\frac{T}{\alpha_2}(L_0^{(2)}+\bar{L}_0^{(2)})}
\bigl[\cosh\Y_{123}\bigr]^{ij}|V^{\rm f}\rangle_{123}
\bigl[\cosh\Y_{456}\bigr]^{kl}|V^{\rm f}\rangle_{456}.
\label{RVVf}
\end{align}
The superscript ``f'' denotes that we pick up the fermionic sector.
Comparing with the previous case of \eqref{RVV}, we note that in the
computation of \eqref{RVVb} and \eqref{RVVf} we have to deal with the
prefactors $Z^i$ or $[\cosh\Y]^{ij}$.

Since both the reflector and the interaction vertex basically take the
Gaussian form, we will utilize the Gaussian convolution formula in our
computation.
The bosonic case of the Gaussian convolution formula is well known and
can be found, for example, in \cite{KMT}.
For the fermionic oscillators $S_m$, $S^{\dagger}_n$ satisfying
$\{S_m,S^{\dagger}_n\}=\delta_{m,n}$, the formula reads
\begin{align}
&\langle 0|\exp\biggl({1\over 2}\bs{S}^{\rm T}M\bs{S}
+\bs{k}^{\rm T}\bs{S}\biggr)
\exp\biggl({1\over 2}\bs{S}^{\dagger{\rm T}}N\bs{S}^\dagger
+\bs{l}^{\rm T}\bs{S}^\dagger\biggr)|0\rangle\nonumber\\
&=\sqrt{\det(1+MN)}
\exp\biggl({1\over 2}\bs{k}^{\rm T}N\frac{1}{1+MN}\bs{k}
+{1\over 2}\bs{l}^{\rm T}\frac{1}{1+MN}M\bs{l}
+\bs{l}^{\rm T}\frac{1}{1+MN}\bs{k}\biggr).
\label{gaussian}
\end{align}

To deal with the bosonic prefactor $Z_{123}^iZ_{456}^k$ in \eqref{RVVb},
we need to introduce the source term
$e^{\beta_{123}^iZ_{123}^i+\beta_{456}^kZ_{456}^k}$, take the derivative
with respect to $\beta$ and finally set $\beta=0$.
However, the fermionic prefactor
$[\cosh\Y_{123}]^{ij}[\cosh\Y_{456}]^{kl}$ in \eqref{RVVf} looks much
more complicated than the bosonic one $Z_{123}^iZ_{456}^k$.
Fortunately, using the Fourier transformation of the prefactor
\eqref{fourier1} we can compute the contractions with the simple source
term $e^{\frac{2}{\alpha_{123}}(\phi_{123}Y_{123}-\phi_{456}Y_{456})}$
as in the bosonic case and perform the $\phi$ integration afterwards.
The surprise is that, although the right hand side of \eqref{gaussian}
consists of the bilinear terms of the sources $\bs{k}$ and $\bs{l}$,
after the computation we find that the bilinear terms of $\phi$ do
not appear in the final result.
Therefore, in the $\phi$ integration we can apply \eqref{fourier1} once
again and write down the answer explicitly.

The contents of the present paper are as follows.
In the next section, we recapitulate some ingredients of LCSFT, which
will be necessary in our computation.
After the review we proceed to the computation of the contractions of
the string interaction vertices which correspond to the OPEs.
We first compute the contractions of the tree diagram in Section 3 in
the ordering of the bosonic sector and the fermionic sector.
Then we turn to the computation of the one-loop diagram in the same
ordering in Section 4.
Finally we conclude our paper with some comments.
Appendix A is devoted to our new notations of the prefactors.
In Appendix B we collect several relations of the Neumann coefficient
matrices and prove some preliminary formulas for Appendix C.
In Appendix C we evaluate the small intermediate time behavior of some
Neumann coefficient matrix products.
In Appendix D we collect some formulas for the gamma matrices.
The results in the appendices are necessary at each stage of our diagram
computation.

\section{LCSFT}
In this section we would like to recapitulate some ingredients of LCSFT,
which are necessary in the computation of following sections.
The first order interaction terms are expressed by the three-string Fock
space and given by \cite{GSB}
\begin{align}
|H_1\rangle_{123}&=Z^i\bar Z^j\bigl[\cosh\Y\bigr]^{ij}
|V\rangle_{123},\\
|Q_1^{\dot a}\rangle_{123}
&=\sqrt{-\alpha_{123}}\bar{Z}^i\bigl[\sinh\Y\bigr]^{\dot{a}i}
|V\rangle_{123},\\
|\tilde Q_1^{\dot a}\rangle_{123}
&=i\sqrt{-\alpha_{123}}Z^i\bigl[\sinh\Y\bigr]^{i\dot{a}}
|V\rangle_{123}.
\end{align}
Here the kinematical interaction vertex $|V\rangle_{123}
=|V^{\rm b}(1_{\alpha_1},2_{\alpha_2},3_{\alpha_3})\rangle
|V^{\rm f}(1_{\alpha_1},2_{\alpha_2},3_{\alpha_3})\rangle$,
determined by the overlapping conditions of strings, is given
by
\begin{align}
&|V^{\rm b}(1_{\alpha_1},2_{\alpha_2},3_{\alpha_3})\rangle
=\int\delta^{\rm b}(1,2,3)
e^{E^{\rm b}+\bar{E}^{\rm b}}
|p_1\rangle_1|p_2\rangle_2|p_3\rangle_3,\\
&|V^{\rm f}(1_{\alpha_1},2_{\alpha_2},3_{\alpha_3})\rangle
=\int\delta^{\rm f}(1,2,3)e^{E^{\rm f}}
|\lambda_1\rangle_1|\lambda_2\rangle_2|\lambda_3\rangle_3,
\end{align}
with the zero-mode delta functions being
\begin{align}
&\int\delta^{\rm b}(1,2,3)=\int
\frac{d^8p_1}{(2\pi)^8}\frac{d^8p_2}{(2\pi)^8}\frac{d^8p_3}{(2\pi)^8}
(2\pi)^8\delta^8(p_1+p_2+p_3),\\
&\int\delta^{\rm f}(1,2,3)=\int d^8\lambda_1d^8\lambda_2d^8\lambda_3
\delta^8(\lambda_1+\lambda_2+\lambda_3),
\end{align}
and the oscillator bilinears $E^{\rm b}$ (and its anti-holomorphic
cousin $\bar{E}^{\rm b}$) and $E^{\rm f}$ given by
\begin{align}
E^{\rm b}&=\frac{1}{2}
\bs{a}^{(12)\dagger{\rm T}}N^{12,12}\bs{a}^{(12)\dagger}
+\bs{a}^{(12)\dagger{\rm T}}N^{12,3}\bs{a}^{(3)\dagger}
+\frac{1}{2}
\bs{a}^{(3)\dagger{\rm T}}N^{3,3}\bs{a}^{(3)\dagger}\nonumber\\
&\qquad+\bigl(\bs{N}^{12{\rm T}}\bs{a}^{(12)\dagger}
+\bs{N}^{3{\rm T}}\bs{a}^{(3)\dagger}\bigr){\mathbb P}_{123}
-\frac{\tau_0}{2\alpha_{123}}{\mathbb P}_{123}^2,\\
E^{\rm f}&=\frac{1}{2}
\bs{S}^{(12)\dagger{\rm T}}[\hat{N}]^{12,12}\bs{S}^{(12)\dagger}
+\bs{S}^{(12)\dagger{\rm T}}[\hat{N}]^{12,3}\bs{S}^{(3)\dagger}
+\frac{1}{2}
\bs{S}^{(3)\dagger{\rm T}}[\hat{N}]^{3,3}\bs{S}^{(3)\dagger}
\nonumber\\&\qquad
-\sqrt{2}\Lambda_{123}
\bigl(\hat{\bs{N}}{}^{12{\rm T}}\bs{S}^{{\rm II}(12)\dagger}
+\hat{\bs{N}}{}^{3{\rm T}}\bs{S}^{{\rm II}(3)\dagger}\bigr).
\end{align}
The explicit forms of the building blocks of the prefactors $Z^i$ and
$Y^a$, defined in \eqref{Z} and \eqref{Y}, are given by
\begin{align}
&Z^i={\mathbb P}_{123}^i
-\alpha_{123}\sum_{r=1}^3\sum_{n=1}^\infty
\frac{n}{\alpha_r}N_n^ra^{(r)i}_{-n},\\
&Y^a=\Lambda^a_{123}
-\frac{\alpha_{123}}{\sqrt{2}}\sum_{r=1}^3\sum_{n=1}^\infty
\sqrt{\frac{n}{\alpha_r}}N_n^rS^{{\rm I}(r)a}_{-n}.
\end{align}

Our notation for the bosonic part is exactly the same as those in our
previous work \cite{KMT}.
Hence, we shall only explain our notation for the fermionic part in
detail.
We define the zero mode $\Lambda_{123}^a$ to be 
$\Lambda_{123}^a=\alpha_1\lambda_2^a-\alpha_2\lambda_1^a$, and adopt the
matrix notation for the infinite nonzero modes of the fermionic
oscillators $S^{{\rm I}(r)a}_m$ and $S^{{\rm II}(r)a}_m$ satisfying
\begin{align}
\{S^{{\rm I}(r)a}_m,S^{{\rm I}(s)b}_n\}
=\delta_{m+n,0}\delta^{r,s}\delta^{a,b},\quad
\{S^{{\rm II}(r)a}_m,S^{{\rm II}(s)b}_n\}
=\delta_{m+n,0}\delta^{r,s}\delta^{a,b},\quad
\{S^{{\rm I}(r)a}_m,S^{{\rm II}(s)b}_n\}=0,
\end{align}
and
$S^{{\rm I}(r)a}_m|\lambda_r\rangle_r
=S^{{\rm II}(r)a}_m|\lambda_r\rangle_r=0$ ($m>0$)
on the vacuum ket-state $|\lambda_r\rangle_r$.
We first rewrite the fermionic oscillators $S^{{\rm I}(r)a}_m$ and
$S^{{\rm II}(r)a}_m$ into the vector forms $(m>0)$ 
\begin{align}
(\bs{S}^{{\rm I}(r)})_m=S^{{\rm I}(r)}_m,\quad
(\bs{S}^{{\rm II}(r)})_m=S^{{\rm II}(r)}_m,\quad
(\bs{S}^{{\rm I}(r)\dagger})_m=S^{{\rm I}(r)}_{-m},\quad
(\bs{S}^{{\rm II}(r)\dagger})_m=S^{{\rm II}(r)}_{-m}.
\end{align}
Then we combine the fermionic oscillators of the incoming/outgoing
strings as
\begin{align}
\bs{S}^{{\rm I}(12){\rm T}}=\begin{pmatrix}
\bs{S}^{{\rm I}(1){\rm T}}&\bs{S}^{{\rm I}(2){\rm T}}
\end{pmatrix},\quad
\bs{S}^{{\rm II}(12){\rm T}}=\begin{pmatrix}
\bs{S}^{{\rm II}(1){\rm T}}&\bs{S}^{{\rm II}(2){\rm T}}
\end{pmatrix},
\end{align}
and finally pair up the oscillators with the index I and the
oscillators with the index II:
\begin{align}
\bs{S}^{(12){\rm T}}=\begin{pmatrix}
\bs{S}^{{\rm I}(12){\rm T}}&\bs{S}^{{\rm II}(12){\rm T}}
\end{pmatrix},\quad
\bs{S}^{(3){\rm T}}=\begin{pmatrix}
\bs{S}^{{\rm I}(3){\rm T}}&\bs{S}^{{\rm II}(3){\rm T}}
\end{pmatrix}.
\end{align}

Correspondingly, we repeat the same manipulation for the Neumann
coefficient matrices.
The Neumann coefficient matrices for the fermionic oscillators 
$\hat N^{r,s}$, $\hat{\bs{N}}{}^{r}$ are constructed out of those for
the bosonic oscillators $N^{r,s}$, $\bs{N}^r$ by
\begin{align}
\hat N^{r,s}=(C/\alpha_r)^{\frac{1}{2}}
N^{r,s}(C/\alpha_s)^{-\frac{1}{2}},\quad
\hat{\bs{N}}{}^{r}=(C/\alpha_r)^{\frac{1}{2}}\bs{N}^r.
\end{align}
We combine the Neumann coefficient matrices of the incoming/outgoing
strings by
\begin{align}
\hat{N}^{12,12}=\begin{pmatrix}
\hat{N}^{1,1}&\hat{N}^{1,2}\\\hat{N}^{2,1}&\hat{N}^{2,2}
\end{pmatrix},\quad
\hat{N}^{3,12}=\begin{pmatrix}
\hat{N}^{3,1}&\hat{N}^{3,2}\end{pmatrix},\quad
\hat{N}^{12,3}=\begin{pmatrix}
\hat{N}^{1,3}\\\hat{N}^{2,3}\end{pmatrix},\quad
\hat{\bs{N}}{}^{12}=\begin{pmatrix}
\hat{\bs{N}}{}^1\\\hat{\bs{N}}{}^2\end{pmatrix},
\end{align}
and pair the Neumann coefficient matrices for the fermionic oscillators
with the index I and those for the fermionic oscillators with the index
II:
\begin{align}
[\hat{N}]^{12,12}
\!\!=\!\begin{pmatrix}0&\!\!\!-\hat N^{12,12{\rm T}}\\
\hat N^{12,12}&\!\!\!0\end{pmatrix},\quad
[\hat{N}]^{12,3}
\!\!=\!\begin{pmatrix}0&\!\!\!-\hat N^{3,12{\rm T}}\\
\hat N^{12,3}&\!\!\!0\end{pmatrix},\quad
[\hat{N}]^{3,3}
\!\!=\!\begin{pmatrix}0&\!\!\!-\hat N^{3,3{\rm T}}\\
\hat N^{3,3}&\!\!\!0\end{pmatrix}.
\label{hatN}
\end{align}

Using the matrix notation, the fermionic part of the reflector
$\langle R(3,6)|=\langle R^{\rm b}(3,6)|\langle R^{\rm f}(3,6)|$
is expressed as
\begin{align}
\langle R^{\rm f}(3,6)|=\int\delta^{\rm f}(3,6)
{}_3\langle\lambda_3|{}_6\langle\lambda_6|
\exp\bigg(\frac{1}{2}\bs{S}^{(36){\rm T}}R\bs{S}^{(36)}\bigg),
\end{align}
with the vacuum bra-state satisfying
${}_r\langle\lambda_r|S^{{\rm I}(r)}_{-m}
={}_r\langle\lambda_r|S^{{\rm II}(r)}_{-m}=0$
and $\langle\lambda|\lambda'\rangle=\delta^8(\lambda-\lambda')$.
Here the fermionic oscillators are defined as
$\bs{S}^{(36){\rm T}}=(\bs{S}^{(3){\rm T}}\quad\bs{S}^{(6){\rm T}})$
and the matrix $R$ is given by
\begin{align}
R=\begin{pmatrix}0&i[\Sigma^1]\\-i[\Sigma^1]&0\end{pmatrix},\quad
[\Sigma^1]=\begin{pmatrix}0&1\\1&0\end{pmatrix},
\label{Sigma1}
\end{align}
where the matrices $1$ and $0$ in $[\Sigma^1]$ denote the
infinite-dimensional unity matrix and the infinite-dimensional zero
matrix respectively and are the Neumann coefficient matrices for the
oscillators with the indices I and II.

\section{Tree Amplitude}
After the recapitulation of LCSFT in the previous section, we would
like to proceed to realize various OPEs using the string interaction
vertices.
As explained in the introduction, we have two ways to realize the
OPEs.
This section will devote to the calculation of the four-point tree
diagram with two incoming strings 1 (with length $\alpha_1(>0)$) and 2
(with length $\alpha_2(>0)$), joining and splitting again into two
outgoing strings 4 and 5 of the same length as 1 and 2, respectively.
(See Fig. 1.)

\begin{figure}[htbp]
\begin{center}
\scalebox{0.6}[0.6]{\includegraphics{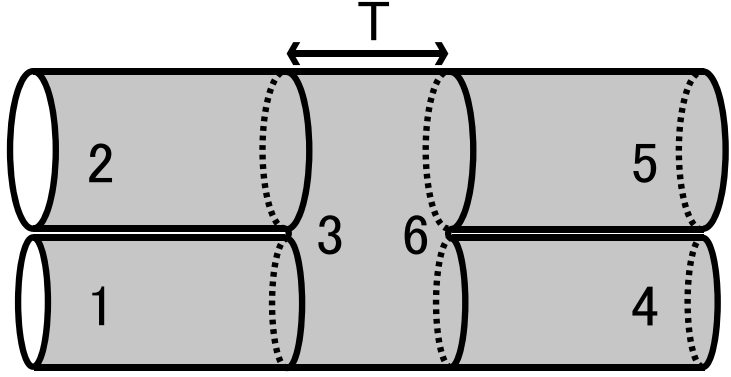}}
\end{center}
\caption{Four-string tree diagram.}
\label{fig:4string}
\end{figure}

\subsection{Bosonic sector}
Let us start with realizing the OPE
\begin{align}
[\tau^i\bar\sigma](z,\bar{z})\cdot[\tau^k\bar\sigma](0)
\sim\frac{\delta^{ik}}{z^2\bar{z}(\log|z|)^4},
\label{twotau}
\end{align}
with the tree diagram.
For this purpose, we shall consider the bosonic effective four-string
interaction vertex
\begin{align}
&|A^{\rm b}(1,2,4,5)\rangle=\langle R^{\rm b}(3,6)|
e^{-{T\over |\alpha_3|}(L_0^{(3)}+\bar{L}_0^{(3)})}\nonumber\\
&\quad\times
Z_{123}^i|V^{\rm b}(1_{\alpha_1},2_{\alpha_2},3_{\alpha_3})\rangle
Z_{456}^k|V^{\rm b}(4_{-\alpha_1},5_{-\alpha_2},6_{-\alpha_3})\rangle.
\end{align}
Comparing the OPE \eqref{twotau} with \eqref{twosigma}, we have an extra
factor of $1/z$.
Therefore, we expect that the extra prefactors $Z_{123}^iZ_{456}^k$ in 
the effective four-string vertex $|A^{\rm b}(1,2,4,5)\rangle$ induce an
extra $1/T$ factor in the limit $T\to+0$.
Here note that we can identify the relative distance in $z$ as the
relative distance in $T$.

For the computation of the effective four-string interaction vertex
$|A^{\rm b}(1,2,4,5)\rangle$, it is sufficient to consider the
generating function
\begin{align}
&|A^{\rm b}_\beta(1,2,4,5)\rangle=\langle R^{\rm b}(3,6)|
e^{-{T\over |\alpha_3|}(L_0^{(3)}+\bar{L}_0^{(3)})}
\nonumber\\
&\quad\times
e^{\beta_{123}^iZ_{123}^i+\beta_{456}^kZ_{456}^k}
|V^{\rm b}(1_{\alpha_1},2_{\alpha_2},3_{\alpha_3})\rangle
|V^{\rm b}(4_{-\alpha_1},5_{-\alpha_2},6_{-\alpha_3})\rangle.
\end{align}
The computation is almost the same as that performed in \cite{KMT}.
The only difference is the source term, but the effect can be easily
absorbed by completing the square.
The result is given as
\begin{align}
&|A^{\rm b}_\beta(1,2,4,5)\rangle
=(\det)^{-8}\int\delta^{\rm b}(1,2,4,5)
e^{F^{\rm b}(1,2,4,5)}\nonumber\\
&\quad\times e^{\beta_{123}{\cal Z}_{123}+\beta_{456}{\cal Z}_{456}
+\frac{1}{2}(\beta_{123}^2+\beta_{456}^2)a_2
+\beta_{123}\beta_{456}b_2}
|p_1\rangle_1|p_2\rangle_2|p_4\rangle_4|p_5\rangle_5,
\label{bosonictree}
\end{align}
where various expressions are
\begin{align}
&\det=\det\Bigl[1-\bigl(e^{-\frac{T}{2|\alpha_3|}C}
N^{3,3}e^{-\frac{T}{2|\alpha_3|}C}\bigr)^2\Bigr]
\sim 2^{-\frac{5}{12}}\mu^{\frac{1}{6}}
\biggl[\frac{T}{|\alpha_{123}|^{1/3}}\biggr]^{\frac{1}{4}},\\
&\qquad\mu=\exp\biggl(-\tau_0\sum_{t=1}^3\frac{1}{\alpha_t}\biggr),
\quad\tau_0=\sum_{t=1}^3\alpha_t\log|\alpha_t|,\\
&\int\delta^{\rm b}(1,2,4,5)
=\int\frac{d^8p_1}{(2\pi)^8}\frac{d^8p_2}{(2\pi)^8}
\frac{d^8p_4}{(2\pi)^8}\frac{d^8p_5}{(2\pi)^8}
(2\pi)^8\delta^8(p_1+p_2+p_4+p_5),\\
&\lim_{T\to +0}F^{\rm b}(1,2,4,5)
=-\bigl(\bs{a}^{(12)\dagger{\rm T}}\bs{a}^{(45)\dagger}
+\bar{\bs{a}}^{(12)\dagger{\rm T}}\bar{\bs{a}}^{(45)\dagger}\bigr)
-(p_1+p_4)^2\lim_{T\to +0}b
\nonumber\\
&\qquad-(p_1+p_4)\alpha_3\bs{N}^{3{\rm T}}(N^{12,3})^{-1}
\bigl(\bs{a}^{(12)\dagger}+\bs{a}^{(45)\dagger}
+\bar{\bs{a}}^{(12)\dagger}+\bar{\bs{a}}^{(45)\dagger}\bigr),
\label{Fb}\\
&b=\alpha_3^2
\bs{N}^{3{\rm T}}\circ\bigl(1-(N^{3,3})^2_\circ\bigr)^{-1}_\circ\bs{N}^3
\sim 2\log\frac{|\alpha_3|}{T},\label{blog}
\end{align}
which have already appeared in \cite{KMT}.
Here we drop the subscript $T$ in $b_T$ of \cite{KMT}, because all the
quantities depend on $T$.
The new effect of the source term is taken care of by
\begin{align}
&{\cal Z}_{123}=(1-a_1)\mathbb{P}_{123}-b_1\mathbb{P}_{456}
-\bs{a}^{\rm T}\bs{a}^{(12)\dagger}
+\bs{b}^{\rm T}\bs{a}^{(45)\dagger},
\label{Z123}\\
&{\cal Z}_{456}=-b_1\mathbb{P}_{123}+(1-a_1)\mathbb{P}_{456}
-\bs{b}^{\rm T}\bs{a}^{(12)\dagger}
+\bs{a}^{\rm T}\bs{a}^{(45)\dagger},
\label{Z456}
\end{align}
with various new quantities defined by
\begin{align}
a_1&=\alpha_{123}\bs{N}^{3{\rm T}}(C/\alpha_3)\circ
\bigl(1-(N^{3,3})^2_\circ\bigr)^{-1}_\circ
 N^{3,3}\circ\bs{N}^3,\\
b_1&=\alpha_{123}\bs{N}^{3{\rm T}}(C/\alpha_3)\circ
\bigl(1-(N^{3,3})^2_\circ\bigr)^{-1}_\circ
\bs{N}^3,\\
a_2&=(\alpha_{123})^2\bs{N}^{3{\rm T}}(C/\alpha_3)\circ
\bigl(1-(N^{3,3})^2_\circ\bigr)^{-1}_\circ
N^{3,3}\circ(C/\alpha_3)\bs{N}^3,\\
b_2&=(\alpha_{123})^2\bs{N}^{3{\rm T}}(C/\alpha_3)\circ
\bigl(1-(N^{3,3})^2_\circ\bigr)^{-1}_\circ
(C/\alpha_3)\bs{N}^3,\\
\bs{a}^{\rm T}
&=\alpha_{123}\Bigl(\bs{N}^{12{\rm T}}(C/\alpha_{12})
+\bs{N}^{3{\rm T}}(C/\alpha_3)\circ
\bigl(1-(N^{3,3})^2_\circ\bigr)^{-1}_\circ
N^{3,3}\circ N^{3,12}\Bigr),\\
\bs{b}^{\rm T}&=\alpha_{123}\bs{N}^{3{\rm T}}(C/\alpha_3)\circ
\bigl(1-(N^{3,3})^2_\circ\bigr)^{-1}_\circ
N^{3,12}.
\end{align}
Here $\circ$ denotes the matrix multiplication with
$e^{-\frac{T}{|\alpha_3|}C}$ inserted and
$(1-(N^{3,3})^2_\circ)^{-1}_\circ$ is defined by
\begin{align}
\bigl(1-(N^{3,3})^2_\circ\bigr)^{-1}_\circ
=1+N^{3,3}\circ N^{3,3}\circ
+N^{3,3}\circ N^{3,3}\circ N^{3,3}\circ N^{3,3}\circ+\cdots.
\end{align}
Note that $C/\alpha_{12}$ is the bookkeeping notation for
\begin{align}
C/\alpha_{12}
=\begin{pmatrix}C/\alpha_1&0\\0&C/\alpha_2\end{pmatrix},
\end{align}
and should not be confused with
$\alpha_{123}=\alpha_1\alpha_2\alpha_3$.
To get back to the effective four-string interaction vertex 
$|A^{\rm b}(1,2,4,5)\rangle$ we need to take the derivative of the
generating function $|A^{\rm b}_\beta(1,2,4,5)\rangle$ and set $\beta=0$
finally:
\begin{align}
|A^{\rm b}(1,2,4,5)\rangle=\frac{\partial}{\partial\beta_{123}^i}
\frac{\partial}{\partial\beta_{456}^k}
|A^{\rm b}_\beta(1,2,4,5)\rangle\biggr|_{\beta=0}
=\bigl(b_2\delta^{ik}+{\cal Z}_{123}^i{\cal Z}_{456}^k\bigr)
|A^{\rm b}_\beta(1,2,4,5)\rangle\biggr|_{\beta=0}.
\end{align}

Let us take the short intermediate time limit $T\to +0$ hereafter to
reproduce the OPE \eqref{twotau}.
Note that the nonzero-mode part of the reflector already appears
correctly in \eqref{Fb} and no extra contributions from the prefactor
\eqref{Z123} and \eqref{Z456} are added to the nonzero-mode part because
of $\lim_{T\to +0}\bs{a}=\lim_{T\to +0}\bs{b}=\bs{0}$ \eqref{aabb}, as
proved in Appendix B.2.
Hence, we shall concentrate on the zero-mode contribution.
Using 
\begin{align}
{\cal Z}_{123}^i\sim-{\cal Z}_{456}^i\sim -b_1\alpha_3(p_1+p_4)^i,
\end{align}
which can be shown with $\lim_{T\to +0}(1-a_1-b_1)=0$ \eqref{aabb}, the
zero-mode part of the effective interaction vertex 
$|A^{\rm b}(1,2,4,5)\rangle$ is given as
\begin{align}
|A^{\rm b}(1,2,4,5)\rangle\biggr|_{0}&\sim\bigl(b_2\delta^{ik}
+(b_1\alpha_3)^2(p_1+p_4)^i(p_1+p_4)^k)\bigr)e^{-b(p_1+p_4)^2}
\delta^8(p_1+p_2+p_4+p_5)\nonumber\\
&=\biggl[\biggl(b_2+\frac{(b_1\alpha_3)^2}{2b}\biggr)\delta^{ik}
+\frac{(b_1\alpha_3)^2}{4b^2}\partial_{p_1^i}\partial_{p_1^k}\biggr]
e^{-b(p_1+p_4)^2}\delta^8(p_1+p_2+p_4+p_5)\nonumber\\
&\sim b_2\delta^{ik}\cdot e^{-b(p_1+p_4)^2}\delta^8(p_2+p_5).
\end{align}
In the last line we have picked up the most singular term using the
short intermediate time behavior of $b$ \eqref{blog}, $b_1$
\eqref{a_1b_1} and $b_2$ \eqref{b_2}.
The behavior of $b_2$ in the short intermediate time limit $T\to+0$ is
$b_2\sim-\alpha_{123}/(2T)$.
Since we have an extra $1/T$ factor compared with the case with no
prefactors, this result is exactly what we have expected from the OPE
\eqref{twotau}.
After all the final result is given as
\begin{align}
|A^{\rm b}(1,2,4,5)\rangle
\sim 2^{-\frac{29}{3}}\pi^{-4}\mu^{-\frac{4}{3}}
|\alpha_{123}|^{\frac{5}{3}}
\frac{\delta^{ik}}{T^3}\biggl(\log\frac{T}{|\alpha_3|}\biggr)^{-4}
|R^{\rm b}(1,4)\rangle|R^{\rm b}(2,5)\rangle.
\label{twotautree}
\end{align}

\subsection{Fermionic sector}
In this subsection we would like to realize the OPE
\begin{align}
\Sigma^i(z)\bar\Sigma^j(\bar z)\cdot\Sigma^k(0)\bar\Sigma^l(0)
\sim\frac{\delta^{ik}\delta^{jl}}{|z|^2},
\label{Sigmaijkl}
\end{align}
in terms of the string interaction vertex.
This OPE corresponds to the fermionic part of the contraction between
two first order Hamiltonians $H_1\cdot H_1$.
Therefore, we shall compute the fermionic effective four-string
interaction vertex
\begin{align}
&|A^{\rm f}(1,2,4,5)\rangle=\langle R^{\rm f}(3,6)|
e^{-{T\over |\alpha_3|}(L_0^{(3)}+\bar{L}_0^{(3)})}\nonumber\\
&\quad\times\bigl[\cosh\Y_{123}\bigr]^{ij}
|V^{\rm f}(1_{\alpha_1},2_{\alpha_2},3_{\alpha_3})\rangle
\bigl[\cosh\Y_{456}\bigr]^{kl}
|V^{\rm f}(4_{-\alpha_1},5_{-\alpha_2},6_{-\alpha_3})\rangle.
\end{align}

Due to the Fourier transformation formula \eqref{fourier1}, we can
evaluate the effective four-string interaction vertex 
$|A^{\rm f}(1,2,4,5)\rangle$ by the generating function
\begin{align}
&|A^{\rm f}_\phi(1,2,4,5)\rangle
=\langle R^{\rm f}(3,6)|
e^{-{T\over|\alpha_3|}(L_0^{(3)}+\bar{L}_0^{(3)})}
\nonumber\\&\quad\times
e^{\frac{2}{\alpha_{123}}(\phi_{123}Y_{123}-\phi_{456}Y_{456})}
|V^{\rm f}(1_{\alpha_1},2_{\alpha_2},3_{\alpha_3})\rangle
|V^{\rm f}(4_{-\alpha_1},5_{-\alpha_2},6_{-\alpha_3})\rangle.
\end{align}
In order to calculate the generating function
$|A^{\rm f}_\phi(1,2,4,5)\rangle$, we first rewrite the reflector and
the interaction vertices into the following expression:
\begin{align}
&\langle R^{\rm f}(3,6)|
e^{-{T\over |\alpha_3|}(L_0^{(3)}+\bar{L}_0^{(3)})}
=\int\delta^{\rm f}(3,6){}_3\langle\lambda_3|{}_6\langle\lambda_6|
\exp\bigg(\frac{1}{2}\bs{S}^{(36){\rm T}}
\tilde{M}\bs{S}^{(36)}\bigg),\\
&e^{\frac{2}{\alpha_{123}}
(\phi_{123}\Lambda_{123}-\phi_{456}\Lambda_{456})}
|V^{\rm f}(1_{\alpha_1},2_{\alpha_2},3_{\alpha_3})\rangle
|V^{\rm f}(4_{-\alpha_1},5_{-\alpha_2},6_{-\alpha_3})\rangle
\nonumber\\&\quad
=\int\delta^{\rm f}(1,2,3)\delta^{\rm f}(4,5,6)
\exp\bigg(\frac{1}{2}\bs{S}^{(36)\dagger{\rm T}}
\tilde{N}\bs{S}^{(36)\dagger}
+\tilde{\bs{l}}{}^{\rm T}\bs{S}^{(36)\dagger}
+\tilde{P}\bigg)
|\lambda_1\rangle_1|\lambda_2\rangle_2
\cdots|\lambda_6\rangle_6,
\end{align}
with $\bs{S}^{(36)}
=\begin{pmatrix}\bs{S}^{(3)}&\bs{S}^{(6)}\end{pmatrix}$,
$\bs{S}^{(1245)}
=\begin{pmatrix}\bs{S}^{(12)}&\bs{S}^{(45)}\end{pmatrix}$ and
\begin{align}
&\tilde{M}=e^{-\frac{T}{|\alpha_3|}C}\begin{pmatrix}
0&i[\Sigma^1]\\-i[\Sigma^1]&0\end{pmatrix},\quad
\tilde{N}=\begin{pmatrix}
[\hat{N}]^{3,3}&0\\0&[\hat{N}]^{3,3}\end{pmatrix},\\
&\tilde{\bs{l}}{}^{\rm T}=\bs{S}^{(1245)\dagger{\rm T}}
\begin{pmatrix}[\hat{N}]^{3,12{\rm T}}&0\\
0&[\hat{N}]^{3,12{\rm T}}\end{pmatrix}
-\sqrt{2}\begin{pmatrix}
\phi_{123}&\Lambda_{123}&i\phi_{456}&i\Lambda_{456}
\end{pmatrix}\hat{\bs{N}}{}^{3{\rm T}},\\
&\tilde{P}=\frac{1}{2}\bs{S}^{(1245)\dagger{\rm T}}
\begin{pmatrix}[\hat{N}]^{12,12}&0\\0&[\hat{N}]^{12,12}\end{pmatrix}
\bs{S}^{(1245)\dagger}
\nonumber\\&\qquad
-\sqrt{2}\begin{pmatrix}
\phi_{123}&\Lambda_{123}&i\phi_{456}&i\Lambda_{456}
\end{pmatrix}\hat{\bs{N}}{}^{12{\rm T}}\bs{S}^{(1245)\dagger}
+\frac{2}{\alpha_{123}}(\phi_{123}\Lambda_{123}-\phi_{456}\Lambda_{456}).
\end{align}
Then the calculation can be easily done with the help of the fermionic
Gaussian convolution formula \eqref{gaussian}.
The result is given as
\begin{align}
&|A^{\rm f}_\phi(1,2,4,5)\rangle=(\det)^8\int\delta^{\rm f}(1,2,4,5)
e^{\frac{2}{\alpha_{123}}
(\phi_{123}{\cal Y}_{123}-\phi_{456}{\cal Y}_{456})}
e^{F^{\rm f}(1,2,4,5)}
|\lambda_1\rangle_1|\lambda_2\rangle_2
|\lambda_4\rangle_4|\lambda_5\rangle_5,
\label{fermionictree}
\end{align}
with various expressions defined as
\begin{align}
&\int\delta^{\rm f}(1,2,4,5)
=\int d^8\lambda_1d^8\lambda_2d^8\lambda_4d^8\lambda_5
\delta^8(\lambda_1+\lambda_2+\lambda_4+\lambda_5),\\
&F^{\rm f}(1,2,4,5)=\frac{1}{2}\bs{S}^{(1245)\dagger{\rm T}}
M\bs{S}^{(1245)\dagger}
+\bs{k}^{\rm T}\bs{S}^{(1245)\dagger},\\
&{\cal Y}_{123}=(1-a_1)\Lambda_{123}
-b_1\Lambda_{456}
-{1\over\sqrt{2}}\bs{S}^{{\rm I}(12)\dagger{\rm T}}
(C/\alpha_{12})^{-\frac{1}{2}}\bs{a}
+{i\over\sqrt{2}}\bs{S}^{{\rm I}(45)\dagger{\rm T}}
(C/\alpha_{12})^{-{1\over 2}}\bs{b},\\
&{\cal Y}_{456}=-b_1\Lambda_{123}+(1-a_1)\Lambda_{456}
+{1\over\sqrt{2}}\bs{S}^{{\rm I}(12)\dagger{\rm T}}
(C/\alpha_{12})^{-{1\over 2}}\bs{b}
-{i\over\sqrt{2}}\bs{S}^{{\rm I}(45)\dagger{\rm T}}
(C/\alpha_{12})^{-{1\over 2}}\bs{a}.
\end{align}
Here the effective Neumann coefficient matrix $M$ and $\bs{k}^{\rm T}$
are given as
\begin{align}
&M=\begin{pmatrix}
[A]&i[B]\\-i[B]&[A]\end{pmatrix},\quad
[A]=\begin{pmatrix}
0&-A^{\rm T}\\A&0\end{pmatrix},\quad
[B]=\begin{pmatrix}
0&B^{\rm T}\\B&0\end{pmatrix},\\
&\bs{k}^{{\rm T}}=-\sqrt{2}\begin{pmatrix}
0&\Lambda_{123}\bs{U}^{\rm T}
-\alpha_3(\lambda_1+\lambda_4)\bs{V}^{\rm T}&
0&i(\Lambda_{456}\bs{U}^{\rm T}
+\alpha_3(\lambda_1+\lambda_4)\bs{V}^{\rm T})
\end{pmatrix},
\end{align}
with the building blocks being
\begin{align}
&A=\hat{N}^{12,12}+\hat{N}^{12,3}\circ\hat{N}^{3,3}\circ
\bigl(1-(\hat{N}^{3,3})^2_\circ\bigr)^{-1}_\circ
\hat{N}^{3,12},\\
&B=\hat{N}^{12,3}\circ
\bigl(1-(\hat{N}^{3,3})^2_\circ\bigr)^{-1}_\circ
\hat{N}^{3,12},\\
&\bs{U}=\hat{\bs{N}}^{12}
+\hat{N}^{12,3}\circ\bigl(1-\hat{N}^{3,3}\bigr)^{-1}_\circ
\hat{\bs{N}}^{3},\\
&\bs{V}=\hat{N}^{12,3}\circ
\bigl(1-(\hat{N}^{3,3})^2_\circ\bigr)^{-1}_\circ
\hat{\bs{N}}^{3}.
\end{align}
What is surprising in this result \eqref{fermionictree} is that although
in the bosonic case we find the linear source term $\beta$ induces the
squared source term $\beta^2$ in the final result \eqref{bosonictree},
the squared source term $\phi^2$ is absent in the current final result
\eqref{fermionictree}.
The reason is that both the Neumann coefficient matrix in the
interaction vertex \eqref{hatN} and that in the reflector
\eqref{Sigma1} connect the oscillators with the index I and the
oscillators with the index II, but in the Fourier transformation
formula \eqref{fourier1} the source $\phi$ only couples to the
oscillators with the index I.
Due to this fact, we can perform the $\phi$ integration without
difficulty.
After performing the inverse Fourier transformation \eqref{fourier1} for
the result of the generating function $|A^{\rm f}_\phi(1,2,4,5)\rangle$
\eqref{fermionictree}, we find the effective four-string interaction
vertex $|A^{\rm f}(1,2,4,5)\rangle$ itself is given by
\begin{align}
&|A^{\rm f}(1,2,4,5)\rangle
=(\det)^8\int\delta^{\rm f}(1,2,4,5)\nonumber\\
&\quad\times\bigl[\cosh\calY_{123}\bigr]^{ij}
\bigl[\cosh\calY_{456}\bigr]^{kl}
e^{F^{\rm f}(1,2,4,5)}
|\lambda_1\rangle_1|\lambda_2\rangle_2
|\lambda_4\rangle_4|\lambda_5\rangle_5.
\end{align}

Note that the kinematical overlapping part in the short intermediate
time limit $T\to +0$ is given as
\begin{align}
&\lim_{T\to +0}A=0,\quad\lim_{T\to +0}B=1,\quad
\lim_{T\to +0}\bs{U}=\bs{0},\\
&\lim_{T\to +0}\bs{V}=\bigl(C/\alpha_{12}\bigr)^{\frac{1}{2}}
\bigl(N^{3,12}\bigr)^{-1}\bs{N}^3
=-\frac{\alpha_3}{2}\bigl(C/\alpha_{12}\bigr)^{\frac{3}{2}}
A^{(12){\rm T}}C^{-1}\bs{B},
\end{align}
which implies
\begin{align}
&\lim_{T\to +0}F^{\rm f}(1,2,4,5)
=i(\bs{S}^{{\rm I}(45)\dagger{\rm T}}\bs{S}^{{\rm II}(12)\dagger}
-\bs{S}^{{\rm I}(12)\dagger{\rm T}}\bs{S}^{{\rm II}(45)\dagger})
\nonumber\\
&\qquad-\sqrt{2}\alpha_3(\lambda_1+\lambda_4)
\lim_{T\to +0}\bs{V}^{\rm T}
(i\bs{S}^{{\rm II}(45)\dagger}-\bs{S}^{{\rm II}(12)\dagger}),
\label{Ff}
\end{align}
while the singular behavior of the prefactors are
\begin{align}
{\cal Y}_{123}^a\sim-{\cal Y}_{456}^a\sim{\cal Y}^a,\quad
{\cal Y}^a=b_1\alpha_3(\lambda_1+\lambda_4)^a,\label{calY}
\end{align}
if we use the short intermediate time behavior of $1-a_1$, $b_1$,
$\bs{a}$ and $\bs{b}$ in \eqref{aabb}.
Since we have normalized $Y$ by $\sqrt{-\alpha_{123}}$ as in
\eqref{Yslash}, the result of \eqref{calY} implies
\begin{align}
\calY_{456}=\frac{\sqrt{2}}{\sqrt{-\alpha_{456}}}
\eta^*{\cal Y}_{456}^a\hat\gamma^a
\sim\frac{\sqrt{2}}{i\sqrt{-\alpha_{123}}}
\eta^*(-{\cal Y}_{123}^a)\hat\gamma^a
=i\calY_{123},
\end{align}
where the phase $i$ induce the effect of the transposition:
\begin{align}
\bigl[\cosh i\calY_{123}\bigr]^{kl}
=\bigl[\cosh\calY_{123}\bigr]^{lk}.
\label{coshiY}
\end{align}

To reproduce the tensor structure of the most singular term in the OPE
\eqref{Sigmaijkl}, let us first rewrite the prefactors into
\begin{align}
\bigl[\cosh\calY_{123}\bigr]^{ij}\bigl[\cosh\calY_{123}\bigr]^{lk}
=\frac{1}{2^4}\sum_{m=0}^8{(-1)^{\frac{1}{2}m(m-1)}\over m!}
\hat{\gamma}^{c_1\cdots c_m}_{ik}
(\cosh\calY_{123}\hat{\gamma}^{c_1\cdots c_m}\cosh\calY_{123})_{lj},
\label{coshcoshfierz}
\end{align}
and then expand the hyperbolic functions into polynomials to study the
singular behavior of each term.
Here in \eqref{coshcoshfierz} we have used the Fierz identity
\begin{align}
M_{AB}N_{CD}=\frac{(-1)^{|M||N|}}{2^4}
\sum_{m=0}^8{(-1)^{{1\over 2}m(m-1)}\over m!}
\hat{\gamma}^{c_1\cdots c_m}_{AD}
(N\hat{\gamma}^{c_1\cdots c_m}M)_{CB},
\label{Fierz}
\end{align}
with $\hat{\gamma}^{c_1\cdots c_m}
=\hat{\gamma}^{[c_1}\cdots\hat{\gamma}^{c_m]}$.
Note that from the index structure of 
$\hat{\gamma}^{c_1\cdots c_m}_{ik}$, the summand of
\eqref{coshcoshfierz} is non-vanishing only when $m$ is even.

Since ${\cal Y}$ is the only singularity, the more ${\cal Y}${}'s we
have, the more singular the expression is.
In order to extract the coefficient of the most singular term ($p+q=8$)
\begin{align}
{\cal Y}^{a_1}\cdots{\cal Y}^{a_p}{\cal Y}^{b_1}\cdots{\cal Y}^{b_q}
=\epsilon^{a_1\cdots a_pb_1\cdots b_q}\delta^8({\cal Y}),\quad
\delta^8({\cal Y})={\cal Y}^1\cdots{\cal Y}^8,
\label{delta8}
\end{align}
we note the formulas \cite{Kugo}
\begin{align}
&\frac{1}{q!}\epsilon^{a_1\cdots a_pb_1\cdots b_q}
\hat\gamma^{b_1\cdots b_q}
=(-1)^{\frac{1}{2}p(p-1)}\hat\gamma^{a_1\cdots a_p}
\hat\gamma_9,
\label{epsilongamma}\\
&\frac{(-1)^{\frac{1}{2}p(p-1)}}{p!}
\hat\gamma^{a_1\cdots a_p}
\hat\gamma^{c_1\cdots c_m}
\hat\gamma^{a_1\cdots a_p}
=d_{p,m}\hat\gamma^{c_1\cdots c_m},
\label{sandwich}
\end{align}
with $\hat{\gamma}_9$ and $d_{p,m}$ defined by 
\begin{align}
\hat{\gamma}_9=\hat{\gamma}^1\cdots\hat{\gamma}^8
=\begin{pmatrix}\delta^{ij}&0\\
0&-\delta_{\dot a\dot b}\end{pmatrix},\quad
(1+x)^{8-m}(1-x)^m=\sum_{p=0}^8(-1)^{pm}d_{p,m}x^p.
\end{align}
There are lots of useful formulas of $d_{p,m}$.
We collect some of them in Appendix D, which are necessary in this
paper.
Using the formulas \eqref{epsilongamma}, \eqref{sandwich} and
\eqref{sumd1}, we find the most singular part of the prefactors is given
as
\begin{align}
\bigl[\cosh\calY_{123}\bigr]^{ij}\bigl[\cosh\calY_{123}\bigr]^{lk}
=16\nu^8
\delta_{ik}\delta_{jl}\delta^8({\cal Y})
+{\cal O}({\cal Y}^6),\quad
\nu=\sqrt{\frac{2}{-\alpha_{123}}}\eta^*.
\label{coshcosh}
\end{align}
This extra fermionic delta function will eliminate the extra term in
$F^{\rm f}(1,2,4,5)$ \eqref{Ff} and turn $|A^{\rm f}(1,2,4,5)\rangle$
into two reflectors
\begin{align}
|A^{\rm f}(1,2,4,5)\rangle
\sim 2^{\frac{26}{3}}\mu^{\frac{4}{3}}|\alpha_{123}|^{-\frac{2}{3}}
\frac{\delta_{ik}\delta_{jl}}{T^2}
|R^{\rm f}(1,4)\rangle|R^{\rm f}(2,5)\rangle,
\label{Sigmaijkltree}
\end{align}
as we have expected from the OPE \eqref{Sigmaijkl}.

\subsection{Other processes}
Other processes corresponding to the OPEs (up to the numerical factor)
\begin{align}
&\Sigma^{\dot{a}}(z)\bar{\Sigma}^i(\bar{z})
\cdot\Sigma^{\dot{b}}(0)\bar{\Sigma}^j(0)
\sim\frac{\delta^{\dot{a}\dot{b}}\delta^{ij}}{|z|^2},\quad
\Sigma^i(z)\bar{\Sigma}^{\dot{a}}(\bar{z})
\cdot\Sigma^j(0)\bar{\Sigma}^{\dot{b}}(0)
\sim\frac{\delta^{ij}\delta^{\dot{a}\dot{b}}}{|z|^2},
\label{other1}\\
&\Sigma^i(z)\bar{\Sigma}^j(\bar{z})
\cdot\Sigma^{\dot{a}}(0)\bar{\Sigma}^k(0)
\sim\frac{\gamma^i_{a\dot{a}}\delta^{jk}}{\sqrt{z}\bar{z}}
\theta^a(0),\quad
\Sigma^i(z)\bar{\Sigma}^j(\bar{z})
\cdot\Sigma^k(0)\bar{\Sigma}^{\dot{a}}(0)
\sim\frac{\delta^{ik}\gamma^j_{a\dot{a}}}{\sqrt{\bar{z}}z}
\bar\theta^a(0),
\label{other2}\\
&\Sigma^{\dot{a}}(z)\bar{\Sigma}^i(\bar{z})
\cdot\Sigma^j(0)\bar{\Sigma}^{\dot{b}}(0)
\sim\frac{\gamma^j_{a\dot{a}}\gamma^i_{b\dot{b}}}{|z|}
\theta^a(0)\bar\theta^b(0),
\label{other3}
\end{align}
can also be evaluated as in the previous subsection.
These OPEs correspond to the fermionic sector of the
$Q_1^{\dot{a}}\cdot Q_1^{\dot{b}}$, 
$\tilde Q_1^{\dot{a}}\cdot\tilde Q_1^{\dot{b}}$, 
$H_1\cdot Q_1^{\dot{a}}$,
$H_1\cdot\tilde Q_1^{\dot{a}}$ and 
$Q_1^{\dot{a}}\cdot\tilde Q_1^{\dot{b}}$ contractions respectively.
Since all of the string interaction vertices have the identical
overlapping part $|V^{\rm f}\rangle$, we would like to concentrate on
the prefactors hereafter.

The tree diagram corresponding to \eqref{other1} can be evaluated
exactly in the same way except that instead of \eqref{coshiY} we use
\begin{align}
\bigl[\sinh i\calY_{123}\bigr]^{\dot{b}j}
=i\bigl[\sinh\calY_{123}\bigr]^{j\dot{b}},\quad
\bigl[\sinh i\calY_{123}\bigr]^{j\dot{b}}
=i\bigl[\sinh\calY_{123}\bigr]^{\dot{b}j},
\end{align}
and instead of \eqref{sumd1} we use \eqref{sumd2}.
Since the contribution of the prefactors is given by
\begin{align}
&\bigl[\sinh\calY_{123}\bigr]^{\dot{a}i}
\bigl[\sinh\calY_{123}\bigr]^{j\dot{b}}
=-16\nu^8
\delta_{ij}\delta_{\dot a\dot b}\delta^8({\cal Y})
+{\cal O}({\cal Y}^6),
\label{sinhsinh1}\\
&\bigl[\sinh\calY_{123}\bigr]^{i\dot{a}}
\bigl[\sinh\calY_{123}\bigr]^{\dot{b}j}
=16\nu^8
\delta_{ij}\delta_{\dot a\dot b}\delta^8({\cal Y})
+{\cal O}({\cal Y}^6),
\label{sinhsinh2}
\end{align}
the effective interaction vertices are computed to be
\begin{align}
&\langle R^{\rm f}(3,6)|
e^{-\frac{T}{|\alpha_3|}(L_0^{(3)}+\bar{L}_0^{(3)})}
\bigl[\sinh\Y_{123}\bigr]^{\dot{a}i}
|V^{\rm f}(1_{\alpha_1},2_{\alpha_2},3_{\alpha_3})\rangle
\bigl[\sinh\Y_{456}\bigr]^{\dot{b}j}
|V^{\rm f}(4_{-\alpha_1},5_{-\alpha_2},6_{-\alpha_3})\rangle\nonumber\\
&\quad\sim
-i2^{\frac{26}{3}}\mu^{\frac{4}{3}}|\alpha_{123}|^{-\frac{2}{3}}
\frac{\delta_{\dot{a}\dot{b}}\delta_{ij}}{T^2}
|R^{\rm f}(1,4)\rangle|R^{\rm f}(2,5)\rangle,
\label{other11tree}\\
&\langle R^{\rm f}(3,6)|
e^{-\frac{T}{|\alpha_3|}(L_0^{(3)}+\bar{L}_0^{(3)})}
\bigl[\sinh\Y_{123}\bigr]^{i\dot{a}}
|V^{\rm f}(1_{\alpha_1},2_{\alpha_2},3_{\alpha_3})\rangle
\bigl[\sinh\Y_{456}\bigr]^{j\dot{b}}
|V^{\rm f}(4_{-\alpha_1},5_{-\alpha_2},6_{-\alpha_3})\rangle\nonumber\\
&\quad\sim 
i2^{\frac{26}{3}}\mu^{\frac{4}{3}}|\alpha_{123}|^{-\frac{2}{3}}
\frac{\delta_{ij}\delta_{\dot{a}\dot{b}}}{T^2}
|R^{\rm f}(1,4)\rangle|R^{\rm f}(2,5)\rangle.
\label{other12tree}
\end{align}

To evaluate the tree diagram corresponding to \eqref{other2} and
\eqref{other3}, a little more effort is required.
Since the most singular term $p+q=8$ vanishes in this case, we have to 
consider terms with $p+q<8$.
First we note that a natural generalization of \eqref{delta8} for
$p+q<8$ is
\begin{align}
{\cal Y}^{a_1}\cdots{\cal Y}^{a_p}{\cal Y}^{b_1}\cdots{\cal Y}^{b_q}
=\frac{(-1)^{\frac{1}{2}r(r-1)}}{r!}
\epsilon^{d_1\cdots d_ra_1\cdots a_pb_1\cdots b_q}
\frac{\partial}{\partial{\cal Y}^{d_1}}\cdots
\frac{\partial}{\partial{\cal Y}^{d_r}}\delta^8({\cal Y}),
\label{subsingular}
\end{align}
if we define $r=8-(p+q)$.
Using \eqref{subsingular} and \eqref{epsilongamma}, we find two
expressions for each term of the polynomial expansion of the hyperbolic
functions:
\begin{align}
\frac{1}{p!q!}({\cal Y}^a\gamma^a)^p
\hat\gamma^{c_1\cdots c_m}({\cal Y}^b\gamma^b)^q
&=\frac{1}{r!}
G^{c_1\cdots c_m,d_1\cdots d_r}_p
\hat\gamma_9
\frac{\partial}{\partial{\cal Y}^{d_1}}\cdots
\frac{\partial}{\partial{\cal Y}^{d_r}}\delta^8({\cal Y})
\nonumber\\
&=\frac{(-1)^m}{r!}
\tilde G^{d_1\cdots d_r,c_1\cdots c_m}_q
\hat\gamma_9
\frac{\partial}{\partial{\cal Y}^{d_1}}\cdots
\frac{\partial}{\partial{\cal Y}^{d_r}}\delta^8({\cal Y}),
\label{GG}
\end{align}
with $G^{c_1\cdots c_m,d_1\cdots d_r}_p$ and
$\tilde G^{d_1\cdots d_r,c_1\cdots c_m}_q$ defined as
\begin{align}
&G^{c_1\cdots c_m,d_1\cdots d_r}_p
=\frac{(-1)^{\frac{1}{2}p(p-1)}}{p!}
\hat\gamma^{a_1\cdots a_p}
\hat\gamma^{c_1\cdots c_m}
\hat\gamma^{a_1\cdots a_pd_1\cdots d_r},\\
&\tilde G^{d_1\cdots d_r,c_1\cdots c_m}_q
=\frac{(-1)^{\frac{1}{2}q(q-1)}}{q!}
\hat\gamma^{d_1\cdots d_rb_1\cdots b_q}
\hat\gamma^{c_1\cdots c_m}
\hat\gamma^{b_1\cdots b_q}.
\end{align}

For the computation of the tree diagram corresponding to \eqref{other2},
we consider the case $p+q=7$ or $r=1$.
For this purpose, $G^{c_1\cdots c_m,d}_p$ is evaluated in Appendix D:
\begin{align}
G^{c_1\cdots c_m,d}_p
=\biggl(\sum_{\stackrel{0\le p_0\le p}{p_0\equiv p\mod 2}}d_{p_0,m}\biggr)
\hat\gamma^{c_1\cdots c_m}\hat\gamma^d
-\biggl(\sum_{\stackrel{0\le p_0\le p}{p_0\equiv p-1\mod 2}}d_{p_0,m}\biggr)
\hat\gamma^d\hat\gamma^{c_1\cdots c_m}.
\label{Gvalue}
\end{align}
Using \eqref{sumd3} and \eqref{sumd4}, we find the contribution of the
prefactors is given by
\begin{align}
&\bigl[\cosh\calY_{123}\bigr]^{ij}
\bigl[\sinh\calY_{123}\bigr]^{k\dot{a}}
=-8\nu^7
\delta_{jk}\gamma^i_{a\dot a}
{\partial\over\partial{\cal Y}^a}\delta^8({\cal Y})
+{\cal O}({\cal Y}^5),\label{coshsinh1}\\
&\bigl[\cosh\calY_{123}\bigr]^{ij}
\bigl[\sinh\calY_{123}\bigr]^{\dot{a}k}
=8\nu^7
\delta_{ik}\gamma^j_{a\dot a}
{\partial\over\partial{\cal Y}^a}\delta^8({\cal Y})
+{\cal O}({\cal Y}^5).\label{coshsinh2}
\end{align}
To translate these results into the effective interaction vertices, we
need a fermionic expansion formula:
\begin{align}
\left[\frac{\partial}{\partial\xi^a}\delta^8(\xi)\right]
\delta^8(\xi+\eta)e^{\xi^c\zeta^c}
=\left[\frac{\partial}{\partial\xi}
-\frac{\partial}{\partial\eta}-\zeta\right]^a
\delta^8(\xi)\delta^8(\eta),
\label{derive}
\end{align}
which can be proved by
\begin{align}
\xi^b{\partial\over\partial\xi^a}\delta^8(\xi)=
\delta^b_a\delta^8(\xi),\quad
\delta^8(\xi+\eta)=\delta^8(\eta)
+\xi^b\frac{\partial}{\partial\eta^b}\delta^8(\eta)
+{\cal O}(\xi^2),\quad
e^{\xi^c\zeta^c}=1+\xi^c\zeta^c+{\cal O}(\xi^2).
\end{align}
After plugging $\xi^a=(\lambda_1+\lambda_4)^a$,
$\eta^a=(\lambda_2+\lambda_5)^a$ 
and $\zeta^a=-\sqrt{2}\alpha_3\lim_{T\to +0}\bs{V}^{\rm T}
(i\bs{S}^{{\rm II}(45)\dagger}-\bs{S}^{{\rm I}(12)\dagger})$ into
\eqref{derive} for our purpose, we find the expression appearing on the
right hand side can be rewritten into
\begin{align}
&\biggl[\frac{\partial}{\partial\lambda_2}
-\frac{\partial}{\partial\lambda_1}
-\sqrt{2}\alpha_3\lim_{T\to +0}
\bs{V}^{\rm T}(i\bs{S}^{{\rm II}(45)\dagger}
-\bs{S}^{{\rm II}(12)\dagger})\biggr]^a
|R^{\rm f}(1,4)\rangle|R^{\rm f}(2,5)\rangle\nonumber\\
&\quad=\bigl[\vartheta^{(2)}(\sigma_{2,\rm int})
-\vartheta^{(1)}(\sigma_{1,\rm int})\bigr]^a
|R^{\rm f}(1,4)\rangle|R^{\rm f}(2,5)\rangle,
\end{align}
with the fermionic coordinate
\begin{align}
\vartheta^{(r)}(\sigma_r)=\vartheta_r
+\sqrt{\frac{2}{\alpha_r}}\sum_{n=1}^\infty
\biggl(\bigl(S_n^{{\rm I}}+S_{-n}^{{\rm II}}\bigr)
\cos\frac{n\sigma_r}{|\alpha_r|}
+\bigl(S_n^{{\rm II}}-S_{-n}^{{\rm I}}\bigr)
\sin\frac{n\sigma_r}{|\alpha_r|}\biggr),
\end{align}
and $\sigma_{1,\rm int}=\pi\alpha_1$, $\sigma_{2,\rm int}=0$.
Therefore, the effective string interaction vertices corresponding to
\eqref{other2} are given as
\begin{align}
&\langle R^{\rm f}(3,6)|
e^{-\frac{T}{|\alpha_3|}(L_0^{(3)}+\bar{L}_0^{(3)})}
\bigl[\cosh\Y_{123}\bigr]^{ij}
|V^{\rm f}(1_{\alpha_1},2_{\alpha_2},3_{\alpha_3})\rangle
\bigl[\sinh\Y_{456}\bigr]^{\dot{a}k}
|V^{\rm f}(4_{-\alpha_1},5_{-\alpha_2},6_{-\alpha_3})\rangle\nonumber\\
&\quad\sim
-\eta^*2^{\frac{20}{3}}\mu^{\frac{4}{3}}|\alpha_{123}|^{-\frac{2}{3}}
\frac{\delta_{jk}\gamma^i_{a\dot{a}}}{T^{3/2}}
\bigl[\vartheta^{(2)}(\sigma_{\rm int})
-\vartheta^{(1)}(\sigma_{\rm int})\bigr]^a
|R^{\rm f}(1,4)\rangle|R^{\rm f}(2,5)\rangle,
\label{other21tree}\\
&\langle R^{\rm f}(3,6)|
e^{-\frac{T}{|\alpha_3|}(L_0^{(3)}+\bar{L}_0^{(3)})}
\bigl[\cosh\Y_{123}\bigr]^{ij}
|V^{\rm f}(1_{\alpha_1},2_{\alpha_2},3_{\alpha_3})\rangle
\bigl[\sinh\Y_{456}\bigr]^{k\dot{a}}
|V^{\rm f}(4_{-\alpha_1},5_{-\alpha_2},6_{-\alpha_3})\rangle\nonumber\\
&\quad\sim
\eta^*2^{\frac{20}{3}}\mu^{\frac{4}{3}}|\alpha_{123}|^{-\frac{2}{3}}
\frac{\delta_{ik}\gamma^j_{a\dot{a}}}{T^{3/2}}
\bigl[\vartheta^{(2)}(\sigma_{\rm int})
-\vartheta^{(1)}(\sigma_{\rm int})\bigr]^a
|R^{\rm f}(1,4)\rangle|R^{\rm f}(2,5)\rangle.
\label{other22tree}
\end{align}
Here we have abbreviated $\sigma_{1,{\rm int}}$ and
$\sigma_{2,{\rm int}}$ as $\sigma_{\rm int}$.

Finally, let us turn to the effective string interaction vertex
corresponding to \eqref{other3}.
We can easily show with the help of \eqref{sumd2} that $p+q\ge 7$ does
not contribute.
Hence, let us consider the case of $p+q=6$.
Though it is not easy to find the value of $G^{c_1\cdots c_m,d_1d_2}_p$
or $\tilde G^{d_1d_2,c_1\cdots c_m}_q$ separately, we find in Appendix D
that the difference can be evaluated as
\begin{align}
G^{c_1\cdots c_m,d_1d_2}_p-\tilde G^{d_1d_2,c_1\cdots c_m}_p
=\biggl(\sum_{\stackrel{0\le p_0\le p}{p_0\equiv p\mod 2}}
d_{p_0,m}\biggr)
[\hat\gamma^{c_1\cdots c_m},\hat\gamma^{d_1d_2}].
\label{Gdiff}
\end{align}
To use this result properly, let us first combine two expressions of
\eqref{GG} into 
\begin{align}
&\frac{1}{p!q!}\bigl(({\cal Y}^a\hat\gamma^a)^p
\hat\gamma^{c_1\cdots c_m}({\cal Y}^b\hat\gamma^b)^q
+({\cal Y}^a\hat\gamma^a)^q
\hat\gamma^{c_1\cdots c_m}({\cal Y}^b\hat\gamma^b)^p\bigr)
\nonumber\\
&\quad=\frac{1}{4}\bigl(G^{c_1\cdots c_m,d_1d_2}_p
+(-1)^m\tilde G^{d_1d_2,c_1\cdots c_m}_p
+G^{c_1\cdots c_m,d_1d_2}_{6-p}
+(-1)^m\tilde G^{d_1d_2,c_1\cdots c_m}_{6-p}\bigr)\hat\gamma_9
\frac{\partial}{\partial{\cal Y}^{d_1}}
\frac{\partial}{\partial{\cal Y}^{d_2}}\delta^8({\cal Y}).
\label{Gpm}
\end{align}
Fortunately, since we only want to consider the case with $m$ being odd
because of the index structure of the gamma matrices in the summand of
\eqref{Fierz}, \eqref{Gpm} reduces to the difference of 
$G^{c_1\cdots c_m,d_1d_2}_p$ and $\tilde G^{d_1d_2,c_1\cdots c_m}_p$.
Therefore, we can apply the result of \eqref{Gdiff} directly to find
\begin{align}
\bigl[\sinh\calY_{123}\bigr]^{\dot{a}i}
\bigl[\sinh\calY_{123}\bigr]^{\dot{b}j}
\sim\nu^6
(\hat\gamma^c)_{\dot{a}j}([\hat\gamma^c,\hat\gamma^{d_1d_2}])_{\dot{b}i}
\frac{\partial}{\partial{\cal Y}^{d_1}}
\frac{\partial}{\partial{\cal Y}^{d_2}}\delta^8({\cal Y}),
\end{align}
with the help of \eqref{sumd5}.
Using the formula for the gamma matrices
$(\hat\gamma^c)_{\dot{a}j}([\hat\gamma^c,\hat\gamma^{d_1d_2}])_{\dot{b}i}
=4\hat\gamma^{[d_1}_{j\dot{a}}\hat\gamma^{d_2]}_{i\dot{b}}$,
we arrive at the result:
\begin{align}
&\bigl[\sinh\calY_{123}\bigr]^{\dot{a}i}
\bigl[\sinh\calY_{123}\bigr]^{\dot{b}j}
=4\nu^6
\gamma^j_{a\dot{a}}\gamma^i_{b\dot{b}}
{\partial\over\partial{\cal Y}^a}
{\partial\over\partial{\cal Y}^b}\delta^8({\cal Y})
+{\cal O}({\cal Y}^4).
\label{sinhsinh}
\end{align}
To translate our result into the effective interaction vertex, we
introduce another fermionic expansion formula similar to \eqref{derive}
\begin{align}
&\left[\frac{\partial}{\partial\xi^a}\frac{\partial}{\partial\xi^b}
\delta^8(\xi)\right]
\delta^8(\xi+\eta)e^{\xi^c\zeta^c}
=\left[\frac{\partial}{\partial\xi}-\frac{\partial}{\partial\eta}
-\zeta\right]^a
\left[\frac{\partial}{\partial\xi}-\frac{\partial}{\partial\eta}
-\zeta\right]^b
\delta^8(\xi)\delta^8(\eta),
\label{derivative}
\end{align}
which can be proved with
\begin{align}
\xi^c\frac{\partial}{\partial\xi^a}\frac{\partial}{\partial\xi^b}
\delta^8(\xi)
=-\delta^c_b\frac{\partial}{\partial\xi^a}\delta^8(\xi)
+\delta^c_a\frac{\partial}{\partial\xi^b}\delta^8(\xi),\quad
\xi^c\xi^d
\frac{\partial}{\partial\xi^a}\frac{\partial}{\partial\xi^b}
\delta^8(\xi)
=-(\delta^c_a\delta^d_b-\delta^c_b\delta^d_a)\delta^8(\xi).
\end{align}
Finally, using \eqref{derivative} the effective string interaction
vertex is expressed as
\begin{align}
&\langle R^{\rm f}(3,6)|
e^{-\frac{T}{|\alpha_3|}(L_0^{(3)}+\bar{L}_0^{(3)})}
\bigl[\sinh\Y_{123}\bigr]^{\dot{a}i}
|V^{\rm f}(1_{\alpha_1},2_{\alpha_2},3_{\alpha_3})\rangle
\bigl[\sinh\Y_{456}\bigr]^{j\dot{b}}
|V^{\rm f}(4_{-\alpha_1},5_{-\alpha_2},6_{-\alpha_3})\rangle
\nonumber\\
&\quad\sim
-2^{\frac{14}{3}}\mu^{\frac{4}{3}}|\alpha_{123}|^{-\frac{2}{3}}
\frac{\gamma^j_{a\dot{a}}\gamma^i_{b\dot{b}}}{T}
\bigl[\vartheta^{(2)}(\sigma_{\rm int})
-\vartheta^{(1)}(\sigma_{\rm int})\bigr]^a
\bigl[\vartheta^{(2)}(\sigma_{\rm int})
-\vartheta^{(1)}(\sigma_{\rm int})\bigr]^b
|R^{\rm f}(1,4)\rangle|R^{\rm f}(2,5)\rangle.
\label{other3tree}
\end{align}
This result exactly takes the form expected from \eqref{other3}.

\section{One-Loop Amplitude}
In the previous section we have computed one realization of the OPE via
the tree diagram.
Here we would like to proceed to the other realization via the one-loop
diagram: the incoming string $6$ splits into two short strings and
join again into the outgoing string $3$.
(See Fig.2.)
Since most of the computations are parallel to the previous section,
we will be short in the presentation and put the prime $P'$ on every
corresponding quantity $P$ in the tree diagram to make the similarity
clear.

\begin{figure}[htbp]
\begin{center}
\scalebox{0.6}[0.6]{\includegraphics{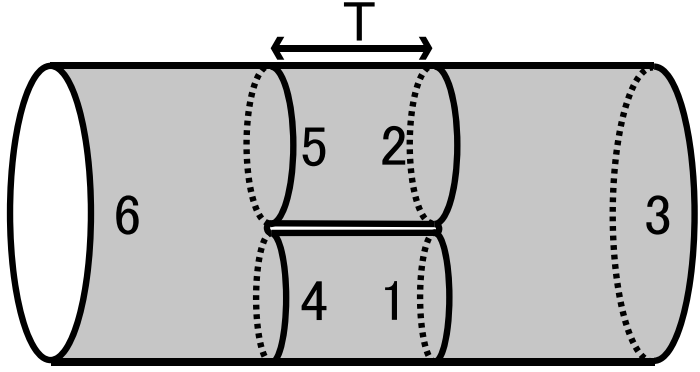}}
\end{center}
\caption{Two-string one-loop diagram.}
\label{fig:loop}
\end{figure}

\subsection{Bosonic sector}
We start with the bosonic sector again.
We would like to compute the effective two-string interaction vertex
\begin{align}
&|A^{\prime\rm b}(3,6)\rangle
=\langle R^{\rm b}(1,4)|\langle R^{\rm b}(2,5)|
e^{-{T\over\alpha_1}(L_0^{(1)}+\bar{L}_0^{(1)})}
e^{-{T\over\alpha_2}(L_0^{(2)}+\bar{L}_0^{(2)})}\nonumber\\
&\quad\times
Z_{123}^i|V^{\rm b}(1_{\alpha_1},2_{\alpha_2},3_{\alpha_3})\rangle
Z_{456}^k|V^{\rm b}(4_{-\alpha_1},5_{-\alpha_2},6_{-\alpha_3})\rangle,
\end{align}
to see whether the prefactor $Z_{123}^iZ_{456}^k$ gives the extra
contribution of $1/T$.
As in the previous section, let us first consider the generating
function
\begin{align}
&|A^{\prime\rm b}_\beta(3,6)\rangle
=\langle R^{\rm b}(1,4)|\langle R^{\rm b}(2,5)|
e^{-{T\over\alpha_1}(L_0^{(1)}+\bar{L}_0^{(1)})}
e^{-{T\over\alpha_2}(L_0^{(2)}+\bar{L}_0^{(2)})}\nonumber\\
&\quad\times e^{\beta_{123}^iZ_{123}^i+\beta_{456}^kZ_{456}^k}
|V^{\rm b}(1_{\alpha_1},2_{\alpha_2},3_{\alpha_3})\rangle
|V^{\rm b}(4_{-\alpha_1},5_{-\alpha_2},6_{-\alpha_3})\rangle.
\end{align}
The result of the generating function is
\begin{align}
&|A^{\prime\rm b}_\beta(3,6)\rangle
=(\det{}')^{-8}\int\frac{d^8p_1}{(2\pi)^8}
\int\delta^{\rm b}(3,6)e^{F^{\prime{\rm b}}(3,6,p_1)}\nonumber\\
&\quad\times
e^{\beta_{123}{\cal Z}'_{123}+\beta_{456}{\cal Z}'_{456}
+\frac{1}{2}(\beta_{123}^2+\beta_{456}^2)a'_2+\beta_{123}\beta_{456}b'_2}
|p_3\rangle_3|p_6\rangle_6,
\end{align}
with various expressions denoting
\begin{align}
&\det{}'=\det\Bigl[1-\bigl(e^{-\frac{T}{2\alpha_{12}}C}
N^{12,12}e^{-\frac{T}{2\alpha_{12}}C}\bigr)^2\Bigr],\\
&\int\delta^{\rm b}(3,6)
=\int\frac{d^8p_3}{(2\pi)^8}\frac{d^8p_6}{(2\pi)^8}
(2\pi)^8\delta^8(p_3+p_6),\\
&\lim_{T\to +0}F(3,6,p_1)
=-\bigl(\bs{a}^{(3)\dagger{\rm T}}\bs{a}^{(6)\dagger}
+\bar{\bs{a}}^{(3)\dagger{\rm T}}\bar{\bs{a}}^{(6)\dagger}\bigr)
+\Bigl(\lim_{T\to +0}c\Bigr)
\biggl[p_1-\frac{\alpha_1}{\alpha_3}p_3\biggr]^2
\nonumber\\&\quad
+\Bigl(\lim_{T\to +0}\bs{C}^{\rm T}\Bigr)
(\bs{a}^{(3)\dagger}-\bs{a}^{(6)\dagger}
+\bar{\bs{a}}^{(3)\dagger}-\bar{\bs{a}}^{(6)\dagger})
\biggl[p_1-\frac{\alpha_1}{\alpha_3}p_3\biggr],\\
&{\cal Z}'_{123}=(1-a'_1-b'_1)\mathbb{P}_{123}
-\bs{a}^{\prime\rm T}\bs{a}^{(3)\dagger}
+\bs{b}^{\prime\rm T}\bs{a}^{(6)\dagger},\\
&{\cal Z}'_{456}=(1-a'_1-b'_1)\mathbb{P}_{123}
-\bs{b}^{\prime\rm T}\bs{a}^{(3)\dagger}
+\bs{a}^{\prime\rm T}\bs{a}^{(6)\dagger}.
\end{align}
Here various quantities are given as
\begin{align}
c&=2\alpha_3^2\Bigl(\frac{T/2-\tau_0}{\alpha_{123}}
+\bs{N}^{12{\rm T}}\circ'\bigl(1-N^{12,12}\bigr)^{-1}_{\circ'}
\bs{N}^{12}\Bigr),\\
\bs{C}&=\alpha_3\Bigl(\bs{N}^3+N^{3,12}\circ'
\bigl(1-N^{12,12}\bigr)^{-1}_{\circ'}\bs{N}^{12}\Bigr),\\
a'_1&=\alpha_{123}\bs{N}^{12{\rm T}}(C/\alpha_{12})\circ'
\bigl(1-(N^{12,12})^2_{\circ'}\bigr)^{-1}_{\circ'}
N^{12,12}\circ'\bs{N}^{12},\\
b'_1&=\alpha_{123}\bs{N}^{12{\rm T}}(C/\alpha_{12})\circ'
\bigl(1-(N^{12,12})^2_{\circ'}\bigr)^{-1}_{\circ'}
\bs{N}^{12},\\
a'_2&=(\alpha_{123})^2\bs{N}^{12{\rm T}}(C/\alpha_{12})\circ'
\bigl(1-(N^{12,12})^2_{\circ'}\bigr)^{-1}_{\circ'}
N^{12,12}\circ'(C/\alpha_{12})\bs{N}^{12},\\
b'_2&=(\alpha_{123})^2\bs{N}^{12{\rm T}}(C/\alpha_{12})\circ'
\bigl(1-(N^{12,12})^2_{\circ'}\bigr)^{-1}_{\circ'}
(C/\alpha_{12})\bs{N}^{12},\\
\bs{a}^{\prime\rm T}&=\alpha_{123}\Bigl(\bs{N}^{3{\rm T}}
(C/\alpha_3)
+\bs{N}^{12{\rm T}}(C/\alpha_{12})\circ'
\bigl(1-(N^{12,12})^2_{\circ'}\bigr)^{-1}_{\circ'}
N^{12,12}\circ'N^{12,3}\Bigr),\\
\bs{b}^{\prime\rm T}&=\alpha_{123}\bs{N}^{12{\rm T}}
(C/\alpha_{12})\circ'
\bigl(1-(N^{12,12})^2_{\circ'}\bigr)^{-1}_{\circ'}N^{12,3},
\end{align}
with $\circ'$ denoting the matrix multiplication with
$e^{-\frac{T}{\alpha_{12}}C}$ inserted.
Again, we dropped the subscript $T$ in $c_T$ and $\bs{C}_T$ from our
previous paper \cite{KMT}.
To consider the effective two-string interaction vertex 
$|A^{\prime\rm b}(3,6)\rangle$, we take the derivative of the generating
function  $|A^{\prime\rm b}_\beta(3,6)\rangle$ as in the previous
section:
\begin{align}
|A^{\prime\rm b}(3,6)\rangle
=\frac{\partial}{\partial\beta_{123}^i}
\frac{\partial}{\partial\beta_{456}^k}
|A^{\prime\rm b}_\beta(3,6)\rangle\biggr|_{\beta=0}
=\bigl(b'_2\delta^{ik}
+{\cal Z}_{123}^{\prime i}{\cal Z}_{456}^{\prime k}\bigr)
|A^{\prime\rm b}_\beta(3,6)\rangle\biggr|_{\beta=0},
\end{align}
where in the last expression the factor $(b'_2\delta^{ik}
+{\cal Z}_{123}^{\prime i}{\cal Z}_{456}^{\prime k})$ should be
interpreted to be in the $p_1$ integration of 
$|A^{\prime\rm b}_\beta(3,6)\rangle$.

Let us consider the short intermediate time limit hereafter.
In \eqref{aabbprime}, we prove that
\begin{align}
\lim_{T\to +0}\bs{a}'=\lim_{T\to +0}\bs{b}'
=\frac{\alpha_{123}}{2}
\biggl[\frac{C}{\alpha_3}\biggr]^{-1}\sum_{t=1}^3
A^{(t)}\biggl[\frac{C}{\alpha_t}\biggr]^2\bs{N}^t,
\end{align}
which, as argued around (C.19) in \cite{GSB}, gives the difference of
the delta functions of physically the same point and vanishes
essentially.
Hence, we can concentrate on the zero-mode part again.
Since we have
\begin{align}
{\cal Z}_{123}^{\prime i}\sim{\cal Z}_{456}^{\prime i}
\sim-2b'_1\alpha_3\Bigl(p_1-\frac{\alpha_1}{\alpha_3}p_3\Bigr)^i,
\end{align}
because of $\lim_{T\to 0}(1-a'_1+b'_1)=0$ \eqref{aabbprime}, the
contribution of the zero-mode part is given by
\begin{align}
&|A^{\prime\rm b}(3,6)\rangle\biggr|_0
\sim\int\frac{d^8p_1}{(2\pi)^8}\bigl(b'_2\delta^{ik}
+4b_1^{\prime 2}\alpha_3^2
(p_1-{\scriptstyle\frac{\alpha_1}{\alpha_3}}p_3)^i
(p_1-{\scriptstyle\frac{\alpha_1}{\alpha_3}}p_3)^k
\bigr)\nonumber\\
&\qquad\times e^{c\left(p_1-\frac{\alpha_1}{\alpha_3}p_3\right)^2
+\bs{C}^{\rm T}\left(\bs{a}^{(3)\dagger}-\bs{a}^{(6)\dagger}
+\bar{\bs{a}}{}^{(3)\dagger}-\bar{\bs{a}}{}^{(6)\dagger}
\right)\left(p_1-\frac{\alpha_1}{\alpha_3}p_3\right)}.
\end{align}
Comparing $b'_2\sim{\cal O}(T^{-1})$ \eqref{b'_2} with 
$b_1^{\prime 2}/c\sim{\cal O}\bigl(T^{-1}(\log T)^{-1}\bigr)$ due to
\eqref{c} and \eqref{a'_1b'_1}, we find that the first term is more
singular.
Because of $\lim_{T\to +0}(\bs{C})_m(\bs{C})_n/c=0$ \cite{KMT} (See also
Appendix C.3.), the leading contribution is given by
\begin{align}
|A^{\prime\rm b}(3,6)\rangle\biggr|_0
\sim b'_2\delta^{ik}\cdot\frac{1}{2^8\pi^4(-c)^4}.
\label{loopzeromode}
\end{align}
The behavior of $b'_2$ \eqref{b'_2} is roughly
$b'_2\sim-\alpha_{123}/(2T)$ in the limit $T\to +0$, so we have found
the expected singular behavior again:
\begin{align}
|A^{\prime{\rm b}}(3,6)\rangle
\sim 2^{-\frac{29}{3}}\pi^{-4}\mu^{-\frac{4}{3}}
|\alpha_{123}|^{\frac{5}{3}}
\frac{\delta^{ik}}{T^3}\biggl(\log\frac{T}{|\alpha_3|}\biggr)^{-4}
|R^{\rm b}(3,6)\rangle.
\label{twotauloop}
\end{align}

\subsection{Fermionic sector}
Let us turn to the fermionic sector of the one-loop diagram.
Here we would like to compute the effective two-string interaction
vertex
\begin{align}
&|A^{\prime\rm f}(3,6)\rangle
=\langle R^{\rm f}(1,4)|\langle R^{\rm f}(2,5)|
e^{-{T\over\alpha_1}(L_0^{(1)}+\bar{L}_0^{(1)})}
e^{-{T\over\alpha_2}(L_0^{(2)}+\bar{L}_0^{(2)})}\nonumber\\
&\quad\times\bigl[\cosh\Y_{123}\bigr]^{ij}
|V^{\rm f}(1_{\alpha_1},2_{\alpha_2},3_{\alpha_3})\rangle
\bigl[\cosh\Y_{456}\bigr]^{kl}
|V^{\rm f}(4_{-\alpha_1},5_{-\alpha_2},6_{-\alpha_3})\rangle.
\end{align}
As in the tree diagram let us consider the generating function first:
\begin{align}
&|A^{\prime{\rm f}}_\phi(3,6)\rangle
=\langle R^{\rm f}(1,4)|\langle R^{\rm f}(2,5)|
e^{-{T\over\alpha_1}(L_0^{(1)}+\bar{L}_0^{(1)})}
e^{-{T\over\alpha_2}(L_0^{(2)}+\bar{L}_0^{(2)})}
\nonumber\\&
\quad\times e^{\frac{2}{\alpha_{123}}
(\phi_{123}Y_{123}-\phi_{456}Y_{456})}
|V^{\rm f}(1_{\alpha_1},2_{\alpha_2},3_{\alpha_3})\rangle
|V^{\rm f}(4_{-\alpha_1},5_{-\alpha_2},6_{-\alpha_3})\rangle.
\end{align}
After applying the Gaussian convolution formula \eqref{gaussian}, we
find
\begin{align}
&|A^{\prime{\rm f}}_\phi(3,6)\rangle=(\det{}')^8\int\delta^{\rm f}(3,6)
\int d^8\lambda_1
e^{\frac{2}{\alpha_{123}}
(\phi_{123}{\cal Y}'_{123}-\phi_{456}{\cal Y}'_{456})}
e^{F^{\prime\rm f}(3,6)}
|\lambda_3\rangle_3|\lambda_6\rangle_6,
\end{align}
with various expressions defined by
\begin{align}
&F^{\prime\rm f}(3,6)
=\frac{1}{2}\bs{S}^{(36)\dagger{\rm T}}M'\bs{S}^{(36)\dagger}
+\bs{k}^{\prime\rm T}\bs{S}^{(36)\dagger},\label{Ffloop}\\
&{\cal Y}'_{123}=(1-a'_1-b'_1)\Lambda_{123}
-\frac{1}{\sqrt{2}}\bs{S}^{{\rm I}(3)\dagger{\rm T}}
(C/\alpha_{3})^{-\frac{1}{2}}\bs{a}'
+\frac{i}{\sqrt{2}}\bs{S}^{{\rm I}(6)\dagger{\rm T}}
(C/\alpha_{3})^{-\frac{1}{2}}\bs{b}',\\
&{\cal Y}'_{456}=(1-a'_1-b'_1)\Lambda_{123}
+{1\over\sqrt{2}}\bs{S}^{{\rm I}(3)\dagger{\rm T}}
(C/\alpha_{3})^{-{1\over 2}}\bs{b}'
-{i\over\sqrt{2}}\bs{S}^{{\rm I}(6)\dagger{\rm T}}
(C/\alpha_{3})^{-{1\over 2}}\bs{a}'.
\end{align}
Here $M'$ and $\bs{k}^{\prime\rm T}$ are given as
\begin{align}
&M'=\begin{pmatrix}
[A']&-i[B']\\i[B']&[A']\end{pmatrix},\quad
[A']=\begin{pmatrix}
0&-A^{\prime\rm T}\\A'&0\end{pmatrix},\quad
[B']=\begin{pmatrix}
0&B^{\prime\rm T}\\B'&0\end{pmatrix},\\
&\bs{k}^{\prime{\rm T}}=-\sqrt{2}\begin{pmatrix}
0&1&0&i\end{pmatrix}
\Lambda_{123}\bs{U}^{\prime\rm T},
\label{kloop}
\end{align}
with the building blocks being
\begin{align}
&A'=\hat{N}^{3,3}+\hat{N}^{3,12}\circ'\hat{N}^{12,12}\circ'
\bigl(1-(\hat{N}^{12,12})^2_{\circ'}\bigr)^{-1}_{\circ'}
\hat{N}^{12,3},\\
&B'=\hat{N}^{3,12}\circ'
\bigl(1-(\hat{N}^{12,12})^2_{\circ'}\bigr)^{-1}_{\circ'}
\hat{N}^{12,3},\\
&\bs{U}'=\hat{\bs{N}}{}^{3}
+\hat{N}^{3,12}\circ'\bigl(1-\hat{N}^{12,12}\bigr)^{-1}_{\circ'}
\hat{\bs{N}}{}^{12}.\label{bsUprime}
\end{align}
Transforming back to the original effective interaction vertex with
\eqref{fourier1}, we find our result is given as
\begin{align}
&|A^{\prime{\rm f}}(3,6)\rangle=(\det{}')^8\int\delta^{\rm f}(3,6)
\int d^8\lambda_1
\bigl[\cosh\calY_{123}'\bigr]^{ij}\bigl[\cosh\calY_{456}'\bigr]^{kl}
e^{F^{\prime\rm f}(3,6)}
|\lambda_3\rangle_3|\lambda_6\rangle_6.
\label{lambda1int}
\end{align}

Note that for $T\to +0$, we have
\begin{align}
\lim_{T\to +0}A'=0,\quad\lim_{T\to +0}B'=1,\quad
\lim_{T\to +0}\bs{U}'=\bs{0},
\end{align}
which implies
\begin{align}
&\lim_{T\to +0}F^{\prime{\rm f}}(3,6)
=i(-\bs{S}^{{\rm I}(6)\dagger{\rm T}}\bs{S}^{{\rm II}(3)\dagger}
+\bs{S}^{{\rm I}(3)\dagger{\rm T}}\bs{S}^{{\rm II}(6)\dagger}).
\end{align}
Also, due to \eqref{aabbprime} which essentially means 
$\lim_{T\to +0}\bs{a}'=\lim_{T\to +0}\bs{b}'=\bs{0}$, we have
\begin{align}
{\cal Y}^{\prime a}_{123}\sim{\cal Y}^{\prime a}_{456}
\sim{\cal Y}^{\prime a},\quad
{\cal Y}^{\prime a}
=-2b'_1\alpha_3
\Bigl(\lambda_1-\frac{\alpha_1}{\alpha_3}\lambda_3\Bigr)^a,
\end{align}
where we have used
$\Lambda_{123}=\Lambda_{456}=\alpha_3\lambda_1-\alpha_1\lambda_3$.
As in the previous section, this relation implies
\begin{align}
\calY'_{456}\sim-i\calY'_{123},\quad
\bigl[\cosh-i\calY'_{123}\bigr]^{kl}
=\bigl[\cosh\calY'_{123}\bigr]^{lk}.
\end{align}
Hence, we can repeat the evaluation with the Fierz identity analogous to
\eqref{coshcosh} and find
\begin{align}
|A^{\prime{\rm f}}(3,6)\rangle
\sim 2^{\frac{26}{3}}\mu^{\frac{4}{3}}|\alpha_{123}|^{-\frac{2}{3}}
\frac{\delta_{ik}\delta_{jl}}
{T^2}|R^{\rm f}(3,6)\rangle,
\label{Sigmaijklloop}
\end{align}
with the use of \eqref{bdet}.
This expression of the effective interaction vertex gives the expected
results from the OPE \eqref{Sigmaijkl}.

\subsection{Other processes}
Similarly, using
\begin{align}
\bigl[\sinh-i\calY'_{123}\bigr]^{\dot{b}j}
=-i\bigl[\sinh\calY'_{123}\bigr]^{j\dot{b}},\quad
\bigl[\sinh-i\calY'_{123}\bigr]^{j\dot{b}}
=-i\bigl[\sinh\calY'_{123}\bigr]^{\dot{b}j},
\end{align}
and the one-loop analogues of \eqref{sinhsinh1} and \eqref{sinhsinh2},
we can also evaluate the effective interaction vertices
\begin{align}
&\langle R^{\rm f}(1,4)|\langle R^{\rm f}(2,5)|
e^{-\frac{T}{\alpha_1}(L_0^{(1)}+\bar{L}_0^{(1)})
-\frac{T}{\alpha_2}(L_0^{(2)}+\bar{L}_0^{(2)})}\nonumber\\
&\qquad\times\bigl[\sinh\Y_{123}\bigr]^{\dot{a}i}
|V^{\rm f}(1_{\alpha_1},2_{\alpha_2},3_{\alpha_3})\rangle
\bigl[\sinh\Y_{456}\bigr]^{\dot{b}j}
|V^{\rm f}(4_{-\alpha_1},5_{-\alpha_2},6_{-\alpha_3})\rangle\nonumber\\
&\quad\sim
i2^{\frac{26}{3}}\mu^{\frac{4}{3}}|\alpha_{123}|^{-\frac{2}{3}}
\frac{\delta_{\dot{a}\dot{b}}\delta_{ij}}{T^2}
|R^{\rm f}(3,6)\rangle,
\label{other11loop}\\
&\langle R^{\rm f}(1,4)|\langle R^{\rm f}(2,5)|
e^{-\frac{T}{\alpha_1}(L_0^{(1)}+\bar{L}_0^{(1)})
-\frac{T}{\alpha_2}(L_0^{(2)}+\bar{L}_0^{(2)})}\nonumber\\
&\qquad\times\bigl[\sinh\Y_{123}\bigr]^{i\dot{a}}
|V^{\rm f}(1_{\alpha_1},2_{\alpha_2},3_{\alpha_3})\rangle
\bigl[\sinh\Y_{456}\bigr]^{j\dot{b}}
|V^{\rm f}(4_{-\alpha_1},5_{-\alpha_2},6_{-\alpha_3})\rangle\nonumber\\
&\quad\sim 
-i2^{\frac{26}{3}}\mu^{\frac{4}{3}}|\alpha_{123}|^{-\frac{2}{3}}
\frac{\delta_{ij}\delta_{\dot{a}\dot{b}}}{T^2}
|R^{\rm f}(3,6)\rangle,
\label{other12loop}
\end{align}
which again give the expected results from the OPEs in \eqref{other1}.

To evaluate the one-loop diagram corresponding to the OPEs in
\eqref{other2} and \eqref{other3}, we need more efforts.
Using the one-loop analogues of \eqref{coshsinh1}, \eqref{coshsinh2} and
\eqref{sinhsinh}, we find that the most singular terms in these cases
do not have eight enough $\Lambda_{123}$'s to survive the $\lambda_1$
integration in an expression similar to \eqref{lambda1int}.
Therefore, we need to take $\Lambda_{123}$ in $\bs{k}'$ \eqref{kloop}
out of the exponential factor $e^{F^{\prime{\rm f}}(3,6)}$
\eqref{Ffloop} to compensate the $\lambda_1$ integration.

An explicit asymptotic expression of $\bs{U}'$ \eqref{bsUprime} is
necessary, since $\Lambda_{123}$ in $\bs{k}'$ \eqref{kloop} always
appears simultaneously with $\bs{U}'$, when we take $\Lambda_{123}$ out
of the exponential part $e^{F^{\prime{\rm f}}(3,6)}$ using the formulas
\begin{align}
\biggl[\frac{\partial}{\partial\xi^a}\delta^8(\xi)\biggr]
e^{\xi^c\zeta^c}
=\biggl[\frac{\partial}{\partial\xi}-\zeta\biggr]^a\delta^8(\xi),\quad
\biggl[\frac{\partial}{\partial\xi^a}
\frac{\partial}{\partial\xi^b}\delta^8(\xi)\biggr]
e^{\xi^c\zeta^c}
=\biggl[\frac{\partial}{\partial\xi}-\zeta\biggr]^a
\biggl[\frac{\partial}{\partial\xi}-\zeta\biggr]^b\delta^8(\xi).
\end{align}
Using $\bs{a}'_0$ and $\bs{b}'_0$ with the definition given by
\eqref{a0loop} and \eqref{b0loop} and the asymptotic expression given by
\eqref{a0loopT} and \eqref{b0loopT}, we find the asymptotic expression
of $\bs{U}'$ is
\begin{align}
\bs{U}'=\sqrt{C/\alpha_3}\frac{\bs{a}'_0+\bs{b}'_0}{\alpha_1\alpha_2}
\sim\sqrt{C/\alpha_3}\frac{\pi^2\alpha_1\alpha_2}
{|\alpha_3|^2\log(T/|\alpha_3|)}C\bs{B},
\end{align}
where we have plugged in the value of $g$ \eqref{eq:gana} which is
determined in Appendix C.4.

Our final result can be summarized by the expression of the fermionic
momentum acting on the reflector
\begin{align}
\bigl[\lambda^{(3)}(\sigma)+\lambda^{(6)}(\sigma)\bigr]|R(3,6)\rangle
=\frac{1}{\sqrt{2\alpha_3\pi}}\sum_{n=1}^\infty
(S^{{\rm II}(3)}_{-n}+iS^{{\rm II}(6)}_{-n})
\sin\frac{n\sigma}{|\alpha_3|}|R(3,6)\rangle,
\end{align}
with the fermionic momentum given by
\begin{align}
\lambda^{(r)}(\sigma)=\frac{1}{2\pi|\alpha_r|}\biggl[
\lambda_r+\sqrt{\frac{\alpha_r}{2}}\sum_{n=1}^\infty
\biggl((S^{{\rm II}(r)}_n+S^{{\rm I}(r)}_{-n})
\cos\frac{n\sigma}{|\alpha_r|}
+(S^{{\rm II}(r)}_{-n}-S^{{\rm I}(r)}_n)
\sin\frac{n\sigma}{|\alpha_r|}\biggr)
\biggr].
\end{align}
The results are given by
\begin{align}
&\langle R^{\rm f}(1,4)|\langle R^{\rm f}(2,5)|
e^{-\frac{T}{\alpha_1}(L_0^{(1)}+\bar{L}_0^{(1)})
-\frac{T}{\alpha_2}(L_0^{(2)}+\bar{L}_0^{(2)})}\nonumber\\
&\qquad\times\bigl[\cosh\Y_{123}\bigr]^{ij}
|V^{\rm f}(1_{\alpha_1},2_{\alpha_2},3_{\alpha_3})\rangle
\bigl[\sinh\Y_{456}\bigr]^{\dot{a}k}
|V^{\rm f}(4_{-\alpha_1},5_{-\alpha_2},6_{-\alpha_3})\rangle\nonumber\\
&\quad\sim
\eta^*2^{\frac{20}{3}}\mu^{\frac{4}{3}}|\alpha_{123}|^{-\frac{2}{3}}
\frac{\delta_{jk}\gamma^i_{a\dot{a}}}{T^{3/2}}
4\pi\bigl[\lambda^{(3)}(\sigma_{3,{\rm int}})
+\lambda^{(6)}(\sigma_{3,{\rm int}})\bigr]^a
|R^{\rm f}(3,6)\rangle,
\label{other21loop}\\
&\langle R^{\rm f}(1,4)|\langle R^{\rm f}(2,5)|
e^{-\frac{T}{\alpha_1}(L_0^{(1)}+\bar{L}_0^{(1)})
-\frac{T}{\alpha_2}(L_0^{(2)}+\bar{L}_0^{(2)})}\nonumber\\
&\qquad\times\bigl[\cosh\Y_{123}\bigr]^{ij}
|V^{\rm f}(1_{\alpha_1},2_{\alpha_2},3_{\alpha_3})\rangle
\bigl[\sinh\Y_{456}\bigr]^{k\dot{a}}
|V^{\rm f}(4_{-\alpha_1},5_{-\alpha_2},6_{-\alpha_3})\rangle\nonumber\\
&\quad\sim
-\eta^*2^{\frac{20}{3}}\mu^{\frac{4}{3}}|\alpha_{123}|^{-\frac{2}{3}}
\frac{\delta_{ik}\gamma^j_{a\dot{a}}}{T^{3/2}}
4\pi\bigl[\lambda^{(3)}(\sigma_{3,{\rm int}})
+\lambda^{(6)}(\sigma_{3,{\rm int}})\bigr]^a
|R^{\rm f}(3,6)\rangle,
\label{other22loop}\\
&\langle R^{\rm f}(1,4)|\langle R^{\rm f}(2,5)|
e^{-\frac{T}{\alpha_1}(L_0^{(1)}+\bar{L}_0^{(1)})
-\frac{T}{\alpha_2}(L_0^{(2)}+\bar{L}_0^{(2)})}\nonumber\\
&\qquad\times\bigl[\sinh\Y_{123}\bigr]^{\dot{a}i}
|V^{\rm f}(1_{\alpha_1},2_{\alpha_2},3_{\alpha_3})\rangle
\bigl[\sinh\Y_{456}\bigr]^{j\dot{b}}
|V^{\rm f}(4_{-\alpha_1},5_{-\alpha_2},6_{-\alpha_3})\rangle\nonumber\\
&\quad\sim
2^{\frac{14}{3}}\mu^{\frac{4}{3}}|\alpha_{123}|^{-\frac{2}{3}}
\frac{\gamma^j_{a\dot{a}}\gamma^i_{b\dot{b}}}{T}
4\pi\bigl[\lambda^{(3)}(\sigma_{3,{\rm int}})
+\lambda^{(6)}(\sigma_{3,{\rm int}})\bigr]^a
4\pi\bigl[\lambda^{(3)}(\sigma_{3,{\rm int}})
+\lambda^{(6)}(\sigma_{3,{\rm int}})\bigr]^b
|R^{\rm f}(3,6)\rangle,
\label{other3loop}
\end{align}
with $\sigma_{3,{\rm int}}=\pi\alpha_2$, which match exactly with the
OPEs \eqref{other2} and \eqref{other3}.

\section{Conclusion}
In this paper, we have completed our previous attempts of realizing
all the OPEs in MST using the interaction vertices in LCSFT.
We have found all the diagrams reproduce the correct OPEs and
established the correspondence between LCSFT and MST.
Especially, we find the OPEs \eqref{twotau}, \eqref{Sigmaijkl},
\eqref{other1}, \eqref{other2}, \eqref{other3} are realized by the tree
diagrams in \eqref{twotautree}, \eqref{Sigmaijkltree},
\eqref{other11tree}, \eqref{other12tree}, \eqref{other21tree},
\eqref{other22tree}, \eqref{other3tree} and the loop diagrams in
\eqref{twotauloop}, \eqref{Sigmaijklloop}, \eqref{other11loop},
\eqref{other12loop}, \eqref{other21loop}, \eqref{other22loop},
\eqref{other3loop}.
It would be interesting to understand the relation between our current
computations and those in \cite{AF} where the Veneziano amplitude was
reproduced from MST.

We have to confess that we do not fully understand why the
holomorphic quantity $\theta^a(z)$ and the anti-holomorphic quantity
$\bar{\theta}^a(\bar{z})$ in \eqref{other2} and \eqref{other3} are
realized as $\vartheta^{(2)}(\sigma_{{\rm int}})
-\vartheta^{(1)}(\sigma_{{\rm int}})$ in the tree diagrams while as
$4\pi\bigl[\lambda^{(3)}(\sigma_{{\rm int}})
+\lambda^{(6)}(\sigma_{{\rm int}})\bigr]$ in the loop diagrams.
Roughly speaking, two sets of fermions are separated as holomorphic
$\theta^a(z)$ and anti-holomorphic $\bar{\theta}^a(\bar{z})$ in MST
while their linear combinations play the role of the fermionic
coordinate $\vartheta(\sigma)$ and the fermionic momentum
$\lambda(\sigma)$ in LCSFT.
But we cannot make the exact correspondence clear.

Aside from the main result, we have several comments.
First of all, the notoriously complicated prefactors in LCSFT are put
into much simpler expressions \eqref{v}--\eqref{tildes}.
As in Appendix A, using these expressions, the supersymmetry algebras
are shown easily.
We hope these expressions will make LCSFT more accessible to non-experts
of the subject.

Secondly, we have performed the computation of the tree and one-loop
diagrams.
At first sight the computation in the fermionic sector seems impossibly
complicated.
Fortunately, since the result of the generating function does not have
the squared term of the source, we can perform the inverse Fourier
transformation without difficulty and write down the result explicitly.
We hope this fact will enable other important calculations in LCSFT.

Having acquired enough information of the first order interaction term,
we would like to turn to the contact terms next.
We wish to report progress in this direction in the near future.

\section*{Acknowledgments}
We would like to thank S.~Dobashi, S.~Fujii, J.~Gomis, H.~Hata,
K.~Murakami, S.~Rey and S.~Teraguchi for valuable discussions and
comments.
Discussions during the KEK Theory Workshop 2006 and the YITP workshop
YITP-W-06-11 are useful.
We are grateful to the organizers of these workshops.
S.\,M. would also like to thank Kyoto University and KEK for hospitality
where part of this work was done.
This work is supported partly by Grant-in-Aid for Young Scientists
(\#18740143) from the Japan Ministry of Education, Culture, Sports,
Science and Technology, partly by Nishina Memorial Foundation,
partly by Inamori Foundation
and partly by funds provided by the US Department
of Energy (DOE) under cooperative research agreement DE-FG02-05ER41360.
The calculations of various Neumann coefficients using {\it Mathematica}
were partly carried out on sushiki at YITP in Kyoto University.

\appendix
\section{Prefactors}
In this section, we would like to recapitulate the prefactors of the
Green-Schwarz-Brink light-cone superstring field theory.
The prefactors were thought to be notoriously complicated.
We would like to show here that we can simplify the expressions of the
prefactors in our new notation.
First of all, we note that due to the triality of $SO(8)$ we can
construct the gamma matrices with the spinor indices
\begin{align}
\hat{\gamma}{}^a
=\begin{pmatrix}0&\hat{\gamma}^a_{i\dot{a}}\\
\hat{\gamma}^a_{\dot{a}i}&0\end{pmatrix},
\end{align}
by the gamma matrices with the vector indices
$\hat{\gamma}^a_{i\dot{a}}=\hat{\gamma}^a_{\dot{a}i}
\equiv\gamma^i_{a\dot{a}}$.
Here we have used $i,j,k,\cdots$ to represent the vector indices,
$a,b,c,\cdots$ to represent the spinor indices and
$\dot{a},\dot{b},\dot{c},\cdots$ to represent the cospinor indices.
The new gamma matrices with the spinor indices satisfy the standard
anti-commutation relations:
\begin{align}
\hat{\gamma}^a_{i\dot{a}}\hat{\gamma}^b_{\dot{a}j}
+\hat{\gamma}^b_{i\dot{a}}\hat{\gamma}^a_{\dot{a}j}
=2\delta^{ab}\delta_{ij},\quad
\hat{\gamma}^a_{\dot{a}i}\hat{\gamma}^b_{i\dot{b}}
+\hat{\gamma}^b_{\dot{a}i}\hat{\gamma}^a_{i\dot{b}}
=2\delta^{ab}\delta_{\dot{a}\dot{b}}.
\label{spinorgamma}
\end{align}
If we define $\Y$ as
\begin{align}
\Y=\sqrt{\frac{2}{-\alpha_{123}}}\eta^*Y^a\hat{\gamma}{}^a
=\begin{pmatrix}0&\Y_{i\dot a}\\
\Y_{\dot{a}i}&0\end{pmatrix},
\end{align}
using the modified gamma matrices $\hat{\gamma}{}^a$, we find the
complicated prefactors of Hamiltonian and two supercharges, $v^{ji}(Y)$,
$s^{i\dot{a}}(Y)$, $\tilde{s}^{i\dot{a}}(Y)$ as well as the auxiliary
quantity $m^{\dot{a}\dot{b}}(Y)$ can be written as
\begin{align}
v^{ji}(Y)=\bigl[\cosh\Y\bigr]^{ij},\quad
&m^{\dot{a}\dot{b}}(Y)=\bigl[\cosh\Y\bigr]^{\dot{a}\dot{b}},\\
s^{i\dot a}(Y)
=\sqrt{-\alpha_{123}}\bigl[\sinh\Y\bigr]^{\dot{a}i},\quad
&\tilde s^{i\dot a}(Y)
=i\sqrt{-\alpha_{123}}\bigl[\sinh\Y\bigr]^{i\dot{a}},
\end{align}
where the indices of the function are consistent because $\cosh$ is an
even function while $\sinh$ is an odd function.

Let us show that the supersymmetry algebra can be proved easily with our
new notation hereafter.
The $Y^a$ derivative and the $Y^a$ multiplication are paired into two
anti-commuting operators $D^a$ and $D^{*a}$.
We shall modify the definition of two anti-commuting operators slightly
by
\begin{align}
D^a=i\sqrt{\frac{-\alpha_{123}}{2}}D_+^a,\quad
D^{*a}=-\sqrt{\frac{-\alpha_{123}}{2}}D_-^a,
\end{align}
with
\begin{align}
D_\pm^a=\sqrt{\frac{-\alpha_{123}}{2}}\frac{1}{\eta^*}
\frac{\partial}{\partial Y^a}
\pm\sqrt{\frac{2}{-\alpha_{123}}}\eta^*Y^a.
\end{align}
Then we can easily find how the operators $D^a_\pm$ act on $\cosh\Y$
and $\sinh\Y$.
Since the derivative $\partial/\partial Y^a$ can act on any $\Y$ in
the polynomial expansion of the hyperbolic functions, we need a
formula to bring $\hat{\gamma}^a$ to the most left side or the most
right side of the expression.
By iterative use of \eqref{spinorgamma}, we find
\begin{align}
&\Y^k\hat{\gamma}^a-(-1)^k\hat{\gamma}^a\Y^k
=(-1)^{k-1}2k\sqrt{\frac{2}{-\alpha_{123}}}\eta^*Y^a\Y^{k-1}
=2k\Y^{k-1}\sqrt{\frac{2}{-\alpha_{123}}}\eta^*Y^a,\\
&\sqrt{\frac{-\alpha_{123}}{2}}\frac{1}{\eta^*}
\frac{\partial}{\partial Y^a}\Y^k
=k\hat{\gamma}^a\Y^{k-1}
-k(k-1)\sqrt{\frac{2}{-\alpha_{123}}}\eta^*Y^a\Y^{k-2}\nonumber\\
&\qquad\qquad\qquad=(-1)^{k-1}\biggl(k\Y^{k-1}\hat{\gamma}^a
-k(k-1)\Y^{k-2}\sqrt{\frac{2}{-\alpha_{123}}}\eta^*Y^a\biggr),
\end{align}
for $k=1,2,\cdots$.
The results of the computation are given as
\begin{align}
&D_+^a\bigl[\cosh\Y\bigr]^{ij}
=\bigl[\hat\gamma^a\sinh\Y\bigr]^{ij},\quad
D_-^a\bigl[\cosh\Y\bigr]^{ij}
=-\bigl[(\sinh\Y)\hat\gamma^a\bigr]^{ij},\\
&D_+^a\bigl[\cosh\Y\bigr]^{\dot{a}\dot{b}}
=\bigl[\hat\gamma^a\sinh\Y\bigr]^{\dot{a}\dot{b}},\quad
D_-^a\bigl[\cosh\Y\bigr]^{\dot{a}\dot{b}}
=-\bigl[(\sinh\Y)\hat\gamma^a\bigr]^{\dot{a}\dot{b}},\\
&D_+^a\bigl[\sinh\Y\bigr]^{\dot{a}i}
=\bigl[\hat\gamma^a\cosh\Y\bigr]^{\dot{a}i},\quad
D_-^a\bigl[\sinh\Y\bigr]^{\dot{a}i}
=\bigl[(\cosh\Y)\hat\gamma^a\bigr]^{\dot{a}i},\\
&D_+^a\bigl[\sinh\Y\bigr]^{i\dot{a}}
=\bigl[\hat\gamma^a\cosh\Y\bigr]^{i\dot{a}},\quad
D_-^a\bigl[\sinh\Y\bigr]^{i\dot{a}}
=\bigl[(\cosh\Y)\hat\gamma^a\bigr]^{i\dot{a}},
\end{align}
which imply
\begin{align}
&D^av^{ij}(Y)=\frac{i}{\sqrt{2}}\gamma^j_{a\dot{a}}s^{i\dot{a}}(Y),\quad
D^{*a}v^{ij}(Y)
=-\frac{i}{\sqrt{2}}\gamma^i_{a\dot{a}}\tilde{s}^{j\dot{a}}(Y),\\
&D^am^{\dot{a}\dot{b}}(Y)
=\frac{1}{\sqrt{2}}\gamma^i_{a\dot{a}}\tilde{s}^{i\dot{b}}(Y),\quad
D^{*a}m^{\dot{a}\dot{b}}(Y)
=\frac{1}{\sqrt{2}}\gamma^i_{a\dot{b}}s^{i\dot{a}}(Y),\\
&D^as^{i\dot{a}}(Y)
=-\frac{i\alpha_{123}}{\sqrt{2}}\gamma^j_{a\dot{a}}v^{ij}(Y),\quad
D^{*a}s^{i\dot{a}}(Y)
=\frac{\alpha_{123}}{\sqrt{2}}
\gamma^i_{a\dot{b}}m^{\dot{a}\dot{b}}(Y),\label{Ds}\\
&D^a\tilde{s}^{i\dot{a}}(Y)
=\frac{\alpha_{123}}{\sqrt{2}}
\gamma^i_{a\dot{b}}m^{\dot{b}\dot{a}}(Y),\quad
D^{*a}\tilde{s}^{i\dot{a}}(Y)
=\frac{i\alpha_{123}}{\sqrt{2}}\gamma^j_{a\dot{a}}v^{ji}(Y).
\label{Dtildes}
\end{align}
Using \eqref{Ds} and \eqref{Dtildes} we can further show
\begin{align}
&\sqrt{2}\alpha_{123}\delta_{\dot{a}\dot{b}}v^{ij}(Y)
=i\gamma^j_{a\dot{a}}D^as^{i\dot{b}}(Y)
+i\gamma^j_{a\dot{b}}D^as^{i\dot{a}}(Y)
=-i\gamma^i_{a\dot{a}}D^{*a}\tilde{s}^{j\dot{b}}(Y)
-i\gamma^i_{a\dot{b}}D^{*a}\tilde{s}^{j\dot{a}}(Y),\\
&\sqrt{2}\alpha_{123}\delta^{ij}m^{\dot{a}\dot{b}}(Y)
=\gamma^j_{a\dot{a}}D^a\tilde{s}^{i\dot{b}}(Y)
+\gamma^i_{a\dot{a}}D^a\tilde{s}^{j\dot{b}}(Y)
=\gamma^j_{a\dot{b}}D^{*a}s^{i\dot{a}}(Y)
+\gamma^i_{a\dot{b}}D^{*a}s^{j\dot{a}}(Y).
\end{align}
All these formulas are sufficient to prove the supersymmetry algebra.

\section{Neumann coefficient matrices}
\subsection{Convention}
We would like to present the definition of various Neumann coefficient
matrices in this appendix, in order to fix the convention used in this 
paper as well as to make preparations for the next appendix.

In \cite{GS} the overlapping condition
was rewritten in terms of the mode expansion and the matrices
$A^{(1)}$, $A^{(2)}$, $\bs{B}$ and $C$ were introduced as
\begin{align}
(A^{(1)})_{mn}&=\sqrt{\frac{n}{m}}\frac{(-1)^m}{\pi\alpha_1}
\!\int_0^{\pi\alpha_1}\!\!\!d\sigma\,
2\cos\frac{n\sigma}{\alpha_1}\cos\frac{m\sigma}{\alpha_3}
=\sqrt{\frac{m}{n}}\frac{(-1)^m}{\pi\alpha_3}
\!\int_0^{\pi\alpha_1}\!\!\!d\sigma\,
2\sin\frac{n\sigma}{\alpha_1}\sin\frac{m\sigma}{\alpha_3},\\
(A^{(2)})_{mn}&=\sqrt{\frac{n}{m}}\frac{(-1)^m}{\pi\alpha_2}
\!\int_{\pi\alpha_1}^{\pi(\alpha_1+\alpha_2)}\!\!\!d\sigma\,
2\cos\frac{n(\sigma-\pi\alpha_1)}{\alpha_2}
\cos\frac{m\sigma}{\alpha_3}\nonumber\\
&=\sqrt{\frac{m}{n}}\frac{(-1)^m}{\pi\alpha_3}
\!\int_{\pi\alpha_1}^{\pi(\alpha_1+\alpha_2)}\!\!\!d\sigma\,
2\sin\frac{n(\sigma-\pi\alpha_1)}{\alpha_2}
\sin\frac{m\sigma}{\alpha_3},\\
(\boldsymbol{B})_m&=\frac{2(-1)^{m+1}}{\sqrt{m}\pi\alpha_1\alpha_2}
\!\int_0^{\pi\alpha_1}\!\!\!d\sigma\,\cos\frac{m\sigma}{\alpha_3}
=\frac{2(-1)^{m}}{\sqrt{m}\pi\alpha_1\alpha_2}
\!\int_{\pi\alpha_1}^{\pi(\alpha_1+\alpha_2)}\!\!\!d\sigma\,
\cos\frac{m\sigma}{\alpha_3},\\
(C)_{mn}&=m\delta_{mn}.
\end{align}
In terms of these matrices, the Neumann coefficient matrices are given 
as
\begin{align}
N^{r,s}=\delta^{rs}-2A^{(r)\rm T}\Gamma^{-1}A^{(s)},\quad
\boldsymbol{N}^r=-A^{(r)\rm T}\Gamma^{-1}\boldsymbol{B},\quad
\Gamma=1+A^{(1)}A^{(1)\rm T}+A^{(2)}A^{(2)\rm T},
\label{Neumann}
\end{align}
if we define $(A^{(3)})_{mn}=\delta_{mn}$ in addition.
It was found in \cite{GS} by explicit computation that these matrices
satisfy the relations ($r,s=1,2$)
\begin{align}
&-\frac{\alpha_r}{\alpha_3}
A^{(r){\rm T}}CA^{(s)}
=\delta_{rs}C,\quad
A^{(r){\rm T}}C\boldsymbol{B}=0,\quad
\frac{1}{2}\alpha_1\alpha_2
\boldsymbol{B}^{{\rm T}}C\boldsymbol{B}=1,\quad
-\frac{\alpha_3}{\alpha_r}
A^{(r){\rm T}}\frac{1}{C}A^{(s)}=\delta_{rs}\frac{1}{C},
\nonumber\\
&\qquad\qquad\sum_{t=1}^3\alpha_tA^{(t)}\frac{1}{C}A^{(t){\rm T}}
=\frac{1}{2}\alpha_1\alpha_2\alpha_3
\boldsymbol{B}\boldsymbol{B}^{{\rm T}},\quad
\sum_{t=1}^3\frac{1}{\alpha_t}A^{(t)}CA^{(t){\rm T}}=0.
\label{unitary}
\end{align}
As was pointed out in \cite{KMT,GMP3}, these relations can simply be
interpreted as the unitarity of the overlapping transformation between
the incoming and outgoing strings where no information is lost.
Due to \eqref{unitary} we can also prove the following relations without
difficulty.
\begin{align}
\sum_{t=1}^3N^{r,t}N^{t,s}=\delta_{r,s},\quad
\sum_{t=1}^3N^{r,t}\bs{N}^t=-\bs{N}^r.
\label{NN}
\end{align}
As in \cite{KMW1}, we adopt
\begin{align}
\bigl(A^{(12)}\bigr)^{-1}
=-\bigl(C/\alpha_{12}\bigr)A^{(12){\rm T}}\bigl(C/\alpha_3\bigr)^{-1},
\label{inverse}
\end{align}
to be the inverse of
$A^{(12)}=\begin{pmatrix}A^{(1)}&A^{(2)}\end{pmatrix}$, 
since we can show that it is a right inverse as well as a left inverse
by applying \eqref{unitary}.

\subsection{Tree diagram formulas}
Here we would like to prove some preliminary formulas
\begin{align}
\lim_{T\to +0}\bs{a}=\lim_{T\to +0}\bs{b}=\bs{0},\quad
\lim_{T\to +0}\bigl((1-a_1)-b_1\bigr)=0,
\label{aabb}
\end{align}
which appear in the main text and will also be necessary in the next
appendix.

Using \eqref{NN} we find that
\begin{align}
&\lim_{T\to +0}\bs{a}^{\rm T}=\alpha_{123}
\biggl(\bs{N}^{12{\rm T}}\frac{C}{\alpha_{12}}
+\bs{N}^{3{\rm T}}\frac{C}{\alpha_3}
N^{3,3}\bigl(N^{12,3}\bigr)^{-1}\biggr),\\
&\lim_{T\to +0}\bs{b}^{\rm T}=\alpha_{123}
\bs{N}^{3{\rm T}}\frac{C}{\alpha_3}\bigl(N^{12,3}\bigr)^{-1}.
\end{align}
With the help of the expression for $N^{r,s}$ in \eqref{Neumann}, we
can put the above two expressions into
\begin{align}
&\lim_{T\to +0}\bs{a}^{\rm T}=\alpha_{123}
\biggl(\bs{N}^{12{\rm T}}\frac{C}{\alpha_{12}}
-\frac{1}{2}\bs{N}^{3{\rm T}}\frac{C}{\alpha_3}
\Gamma\bigl(A^{(12){\rm T}}\bigr)^{-1}
+\bs{N}^{3{\rm T}}\frac{C}{\alpha_3}
\bigl(A^{(12){\rm T}}\bigr)^{-1}\biggr),\label{alim}\\
&\lim_{T\to +0}\bs{b}^{\rm T}
=-\frac{\alpha_{123}}{2}\bs{N}^{3{\rm T}}\frac{C}{\alpha_3}
\Gamma\bigl(A^{(12){\rm T}}\bigr)^{-1}.\label{blim}
\end{align}
If we plug in the expression for $(A^{(12)})^{-1}$
\eqref{inverse} we find the first term and the last term of \eqref{alim}
cancel each other.
Therefore both the expressions \eqref{alim} and \eqref{blim} reduce to
the same form.
Furthermore, if we plug in the expression of $\Gamma$ \eqref{Neumann}
and $(A^{(12)})^{-1}$ \eqref{inverse}, we obtain
\begin{align}
\lim_{T\to +0}\bs{a}^{\rm T}=\lim_{T\to +0}\bs{b}^{\rm T}
=-\frac{\alpha_{123}}{2}\biggl(
-\bs{N}^{3{\rm T}}A^{(12)}\frac{C}{\alpha_{12}}
+\bs{N}^{3{\rm T}}\frac{C}{\alpha_3}A^{(12)}\biggr).
\end{align}
As in \cite{GS}, from the definition of $\Gamma$ \eqref{Neumann}, we can
easily compute $\Gamma CA^{(12)}$ using \eqref{unitary}:
\begin{align}
\Gamma CA^{(12)}=CA^{(12)}-\alpha_3A^{(12)}C/\alpha_{12}.
\end{align}
By multiplying $\bs{B}^{\rm T}\Gamma^{-1}$ from the left, 
we find finally
\begin{align}
\lim_{T\to +0}\bs{a}^{\rm T}=\lim_{T\to +0}\bs{b}^{\rm T}
=\bs{0}^{\rm T}.
\end{align}

For the second formula of \eqref{aabb} let us repeat our formal
computation:
\begin{align}
\lim_{T\to +0}\bigl((1-a_1)-b_1\bigr)
=1-\frac{\alpha_1\alpha_2}{2}\bs{B}^{\rm T}\frac{1}{\Gamma}C\bs{B}.
\end{align}
Since computation of $\Gamma C\bs{B}$ with \eqref{Neumann} and
\eqref{unitary} leads to $\Gamma C\bs{B}=C\bs{B}$, we find
\begin{align}
\bs{B}^{\rm T}\frac{1}{\Gamma}C\bs{B}=\bs{B}^{\rm T}C\bs{B},
\end{align}
which combined with \eqref{unitary} implies the second formula.

\subsection{Loop diagram formulas}
Let us turn to the proof of
\begin{align}
\lim_{T\to +0}\bs{a}^{\prime{\rm T}}
=\lim_{T\to +0}\bs{b}^{\prime{\rm T}}
=\frac{\alpha_{123}}{2}\sum_{t=1}^3\bs{N}^{t{\rm T}}
\biggl[\frac{C}{\alpha_r}\biggr]^2A^{(r){\rm T}}
\biggl[\frac{C}{\alpha_3}\biggr]^{-1}\!\!\!,
\quad\lim_{T\to +0}\bigl((1-a'_1)+b'_1\bigr)=0,
\label{aabbprime}
\end{align}
in this subsection.
The proof is parallel to the previous subsection.
For the first formula, we find both of the expressions reduce to
\begin{align}
\lim_{T\to +0}\bs{a}^{\prime{\rm T}}
=\lim_{T\to +0}\bs{b}^{\prime{\rm T}}
=-\frac{\alpha_{123}}{2}\bs{N}^{12{\rm T}}
\frac{C}{\alpha_{12}}\bigl(A^{(12)}\bigr)^{-1}\Gamma.
\end{align}
Plugging the expression of $\Gamma$ and $(A^{(12)})^{-1}$, we find the
result does not vanish but gives instead \eqref{aabbprime} this time.
The second formula can also be proved similarly.

\section{Small time behavior of the matrix products}
\subsection{Tree diagram formulas}
In this subsection we would like to evaluate $a_1$, $b_1$ and $b_2$,
which is necessary for our analysis of small intermediate time
behavior of the tree diagram amplitude.
Let us define as in \cite{HIKKO2} $(i,j\geq 0)$
\begin{align}
&\bar{a}_{i,j}=\alpha_1\alpha_2\bs{N}^{3{\rm T}}
C^i\circ
N^{3,3}\circ\left(1-(N^{3,3})^2_\circ\right)^{-1}_\circ
C^j\bs{N}^3,\label{aij}\\
&\bar{b}_{i,j}=\alpha_1\alpha_2\bs{N}^{3{\rm T}}
C^i\circ
\left(1-(N^{3,3})^2_\circ\right)^{-1}_\circ C^j
\bs{N}^3.\label{bij}
\end{align}
Then, according to \cite{HIKKO2} we can show that these quantities
satisfy the relations
\begin{align}
&|\alpha_3|{\partial\over\partial T}\log\det=-\bar{a}_{1,1},
\label{logdettree}\\
&|\alpha_3|{\partial\over\partial T}\bar{a}_{i,j}
=\bar{b}_{i,1}\bar{b}_{1,j},\label{atree}\\
&|\alpha_3|{\partial\over\partial T}\bar{b}_{i,j}
=\bar{b}_{i,1}\bar{a}_{1,j}-\bar{b}_{i,j+1},\label{btree}
\end{align}
using the decomposition formula \cite{GS}
\begin{align}
\frac{C}{\alpha_r}N^{r,s}+N^{r,s}\frac{C}{\alpha_s}
=-\alpha_1\alpha_2\alpha_3\frac{C}{\alpha_r}\bs{N}^r
\bs{N}^{s\rm T}\frac{C}{\alpha_s}.
\label{decompose}
\end{align}
Combining with our results from the bosonic case \cite{KMT}
\begin{align}
\det\sim 2^{-\frac{5}{12}}\mu^{\frac{1}{6}}
\biggl[{T\over|\alpha_{123}|^{1/3}}\biggr]^{\frac{1}{4}},\quad
\bar{b}_{0,0}\sim-2\frac{\alpha_1\alpha_2}{\alpha_3^2}
\log{T\over|\alpha_3|},
\end{align}
we have especially
\begin{align}
&1-a_1=1-\bar{a}_{1,0}\sim
\sqrt{\frac{2\alpha_1\alpha_2}{\alpha_3^2}{|\alpha_3|\over T}},\quad
b_1=\bar{b}_{1,0}\sim
\sqrt{\frac{2\alpha_1\alpha_2}{\alpha_3^2}{|\alpha_3|\over T}},
\label{a_1b_1}\\
&b_2=\alpha_1\alpha_2\bar{b}_{1,1}\sim
-{\alpha_{123}\over 2T}.\label{b_2}
\end{align}

\subsection{Loop diagram formulas\label{sec:loop diagram}}
As the Neumann matrix products \eqref{aij} and \eqref{bij} are defined
in \cite{HIKKO2} to analyze the tree diagram amplitude,
let us define
\begin{align}
\bar{a}'_{i,j}&=(\alpha_1\alpha_2)^{\frac{i+j+1}{2}}\alpha_3
\bs{N}^{12{\rm T}}
\left[C/\alpha_{12}\right]^i
\circ'N^{12,12}\circ'
\left(1-(N^{12,12)})^2_{\circ'}\right)^{-1}_{\circ'}
\left[C/\alpha_{12}\right]^j
\bs{N}^{12},\\
\bar{b}'_{i,j}&=(\alpha_1\alpha_2)^{\frac{i+j+1}{2}}\alpha_3
\bs{N}^{12{\rm T}}
\left[C/\alpha_{12}\right]^i\circ'
\left(1-(N^{12,12})^2_{\circ'}\right)^{-1}_{\circ'}
\left[C/\alpha_{12}\right]^j\bs{N}^{12},
\end{align}
for the loop diagram amplitude.
These quantities satisfy the following identities:
\begin{align}
&\sqrt{\alpha_1\alpha_2}{\partial\over\partial T}\log\det{}'
=-\bar{a}'_{1,1},\label{logdetloop}\\
&\sqrt{\alpha_1\alpha_2}{\partial\over\partial T}\bar{a}'_{i,j}
=\bar{b}'_{i,1}\bar{b}'_{1,j},\label{aloop}\\
&\sqrt{\alpha_1\alpha_2}{\partial\over\partial T}\bar{b}'_{i,j}
=\bar{b}'_{i,1}\bar{a}'_{1,j}-\bar{b}'_{i,j+1},\label{bloop}
\end{align}
which imply especially the following relations:
\begin{align}
&\alpha_1\alpha_2{\partial^2\over\partial T^2}\log\det{}'
=-(\bar{b}'_{1,1})^2\\
&\alpha_1\alpha_2{\partial^2\over\partial T^2}
(\bar{a}'_{0,0}+\bar{b}'_{0,0})
=\bar{b}'_{1,1}\left(\bar{b}'_{1,0}-(1-\bar{a}'_{1,0})\right)^2,\\
&\sqrt{\alpha_1\alpha_2}{\partial\over\partial T}
\left(\bar{b}'_{1,0}-(1-\bar{a}'_{1,0})\right)
=\bar{b}'_{1,1}\left(\bar{b}'_{1,0}-(1-\bar{a}'_{1,0})\right).
\end{align}
Combining with the results from our bosonic analysis \cite{KMT},
\begin{align}
&\sqrt{-c}\det{}'\sim 2^{\frac{1}{12}}\mu^{\frac{1}{6}}
\left[{T\over|\alpha_{123}|^{1/3}}
\left(\log{T\over|\alpha_3|}\right)^2\right]^{\frac{1}{4}},
\label{cdet}\\
&c=\frac{\alpha_3(T-2\tau_0)}{\alpha_1\alpha_2}
+\frac{2\alpha_3}{\sqrt{\alpha_1\alpha_2}}
(\bar{a}'_{0,0}+\bar{b}'_{0,0}),
\end{align}
we find that the quantities
$a'_1=\bar{a}'_{1,0}$,
$b'_1=\bar{b}'_{1,0}$ and
$b'_2=\sqrt{\alpha_1\alpha_2}\alpha_3\bar{b}'_{1,1}$
appearing in the main text should satisfy
\begin{align}
&\left(b'_2\over\alpha_{123}\right)^2\sim{1\over 4T^2}
\biggl[\,1+2\biggl(\log\frac{T}{|\alpha_3|}\biggr)^{\!-1}
\!\!\!\!\!+2\biggl(\log\frac{T}{|\alpha_3|}\biggr)^{\!-2}\,\biggr]
+{1\over 2}{\partial^2\over\partial T^2}\log(-c),\\
&{\alpha_1\alpha_2\over\alpha_3}{\partial^2\over\partial T^2}c
=2{b'_2\over\alpha_{123}}\bigl(b'_1-(1-a'_1)\bigr)^2,\\
&{\alpha_{123}\over b'_2}{\partial\over\partial T}
\bigl(b'_1-(1-a'_1)\bigr)=b'_1-(1-a'_1).
\end{align}
Solving the asymptotic behavior by first adopting the ansatz of the
Laurent expansion of $T$ and then correcting by the Laurent expansion of
$\log T$, we find that
\begin{align}
&b'_1-(1-a'_1)\sim\frac{g}{\sqrt{T}}
\biggl(\log{T\over|\alpha_3|}\biggr)^{\!-1},\quad
{b'_2\over\alpha_{123}}\sim-\frac{1}{T}\biggl[\,{1\over 2}
+\biggl(\log{T\over|\alpha_3|}\biggr)^{\!-1}\,\biggr],\quad
\label{b'_2}\\
&{\alpha_1\alpha_2\over\alpha_3}c\sim-g^2
\biggl(\log{T\over|\alpha_3|}\biggr)^{\!-1},
\label{c}
\end{align}
where $g$ is an undetermined constant independent of the intermediate
time $T$.
Moreover, due to \eqref{aabbprime}, we find explicitly
\begin{align}
b'_1\sim\frac{g}{2\sqrt{T}}
\biggl(\log{T\over|\alpha_3|}\biggr)^{-1},\quad
1-a'_1\sim-\frac{g}{2\sqrt{T}}
\biggl(\log{T\over|\alpha_3|}\biggr)^{-1}.
\label{a'_1b'_1}
\end{align}
Note that a combination $2b'_1/\sqrt{-c}$ does not depend on the
undetermined constant $g$.
Combining with \eqref{cdet} we find especially
\begin{align}
\bigl(2b'_1\det{}'\bigr)^8
\sim\frac{2^{\frac{2}{3}}\mu^{\frac{4}{3}}}
{|\alpha_{123}|^{2/3}T^2}
\left(\frac{\alpha_1\alpha_2}{\alpha_3}\right)^4,
\label{bdet}
\end{align}
which appears in the main text.

\subsection{Some Identities}
In this subsection let us make a small digression to clarify several
relations of the Neumann coefficient products.
As a result, among others, we will show
\begin{align}
\lim_{T\to+0}\frac{(\bs{C})_m(\bs{C})_n}{c}=0,
\label{CC/c}
\end{align}
which was conjectured in \cite{KMT} and is also needed in our
computation in \eqref{loopzeromode}.
Our result in this subsection will also enable the evaluation of $g$ in
the next subsection.

We start with proving
\begin{align}
(1-a'_1)^2-b_1^{\prime 2}=1.
\label{a1b1}
\end{align}
Our strategy is basically the same as the derivation of the differential
equations \eqref{logdettree}--\eqref{btree} and
\eqref{logdetloop}--\eqref{bloop}.
First of all let us rewrite $a'_1$ as follows:
\begin{align}
a'_1=\frac{\alpha_{123}}{2}\bs{N}^{12\rm T}
\biggl[\frac{C}{\alpha_{12}}\frac{N^{12,12}}{1-(N^{12,12})^2}
+\frac{N^{12,12}}{1-(N^{12,12})^2}\frac{C}{\alpha_{12}}\biggr]
\bs{N}^{12}.
\label{a1sym}
\end{align}
Here (until \eqref{resum}) note that $\circ'$ in the multiplication
between Neumann matrices is implicit.
We omit it shortly just to simplify our notation.
The key point is to regard the decomposition formula \eqref{decompose}
as an anti-commutation relation between the Neumann coefficient matrix
$N^{12,12}$ and $C/\alpha_{12}$ and move $C/\alpha_{12}$ all the way
from the right to the left.
The quantity in the square bracket of \eqref{a1sym} is given as
($C_{12}=C/\alpha_{12}$)
\begin{align}
\bigl[\cdots\bigr]&=C_{12}N^{12,12}+N^{12,12}C_{12}\nonumber\\
&\qquad+C_{12}(N^{12,12})^3+N^{12,12}C_{12}(N^{12,12})^2
-N^{12,12}C_{12}(N^{12,12})^2-(N^{12,12})^2C_{12}N^{12,12}\nonumber\\
&\qquad\qquad+(N^{12,12})^2C_{12}N^{12,12}+(N^{12,12})^3C_{12}\nonumber\\
&\qquad+\cdots,
\end{align}
which can be resumed into
\begin{align}
\bigl[\cdots\bigr]
&=-\alpha_{123}\biggl\{
\frac{1}{1-(N^{12,12})^2}\frac{C}{\alpha_{12}}\bs{N}^{12}
\bs{N}^{12\rm T}\frac{C}{\alpha_{12}}\frac{1}{1-(N^{12,12})^2}
\nonumber\\&\qquad
-\frac{N^{12,12}}{1-(N^{12,12})^2}\frac{C}{\alpha_{12}}\bs{N}^{12}
\bs{N}^{12\rm T}\frac{C}{\alpha_{12}}\frac{N^{12,12}}{1-(N^{12,12})^2}
\biggr\},
\label{resum}
\end{align}
where we have used the decomposition formula \eqref{decompose}.
This implies that \eqref{a1sym} can be expressed as
\begin{align}
a'_1=\frac{1}{2}\bigl(-(b'_1)^2+(a'_1)^2\bigr),
\end{align}
which is exactly what we want in \eqref{a1b1}.

Similarly, if we further define $\bs{a}'_0$ and $\bs{b}'_0$ as
\begin{align}
&\bs{a}'_0=\alpha_1\alpha_2\Bigl(\bs{N}^3
+N^{3,12}\circ'\bigl(1-(N^{12,12})^2_{\circ'}\bigr)^{-1}_{\circ'}
N^{12,12}\circ'\bs{N}^{12}\Bigr),\label{a0loop}\\
&\bs{b}'_0=\alpha_1\alpha_2
N^{3,12}\circ'\bigl(1-(N^{12,12})^2_{\circ'}\bigr)^{-1}_{\circ'}
\bs{N}^{12},\label{b0loop}
\end{align}
we can prove the following formulas algebraically,
\begin{align}
&\bs{a}'-C\bs{a}'_0=\bs{a}'a'_1-\bs{b}'b'_1,\\
&\bs{b}'+C\bs{b}'_0=-\bs{a}'b'_1+\bs{b}'a'_1.
\end{align}
Using \eqref{a1b1}, we can solve these equations for $\bs{a}'$ and
$\bs{b}'$:
\begin{align}
&\bs{a}'=(1-a'_1)C\bs{a}'_0+b'_1C\bs{b}'_0,\label{veca1}\\
&\bs{b}'=-b'_1C\bs{a}'_0-(1-a'_1)C\bs{b}'_0.
\label{vecb1}
\end{align}

We have found several algebraical formulas thus far.
Let us turn to the proof of the formula \eqref{CC/c}.
Noting $\bs{C}$ can be expressed as 
\begin{align}
\bs{C}=\frac{\alpha_3}{\alpha_1\alpha_2}(\bs{a}'_0+\bs{b}'_0),
\end{align}
let us study the short intermediate time behavior of $\bs{a}'_0$ and
$\bs{b}'_0$.
Our strategy is as follows.
We first consider the derivatives of $\bs{a}'_0$ and $\bs{b}'_0$.
{}From the experience of the previous two subsections, we know roughly
the results should be given by $\bs{a}'$ and $\bs{b}'$, which are
expressed again by $\bs{a}'_0$ and $\bs{b}'_0$ by \eqref{veca1} and
\eqref{vecb1}.
Therefore we can solve the differential equations explicitly.

Similarly to the previous two subsections, we find that
\begin{align}
&\alpha_3\frac{d}{dT}\bs{a}'_0=\bs{b}'b'_1
=-(b'_1)^2C\bs{a}'_0-(1-a'_1)b'_1C\bs{b}'_0,\\
&\alpha_3\frac{d}{dT}\bs{b}'_0=-\bs{b}'(1-a'_1)
=(1-a'_1)b'_1C\bs{a}'_0+(1-a'_1)^2C\bs{b}'_0,
\end{align}
where in the last equations we have used \eqref{veca1} and
\eqref{vecb1}.
Plugging the small intermediate time behavior of $1-a'_1$ and $b'_1$
\eqref{a'_1b'_1}, we find
\begin{align}
&\frac{d}{d\log(T/|\alpha_3|)}\bs{a}'_0
\sim-\frac{g^2}{4\alpha_3\bigl(\log(T/|\alpha_3|)\bigr)^2}
C(\bs{a}'_0-\bs{b}'_0),\label{smallveca0}\\
&\frac{d}{d\log(T/|\alpha_3|)}\bs{b}'_0
\sim-\frac{g^2}{4\alpha_3\bigl(\log(T/|\alpha_3|)\bigr)^2}
C(\bs{a}'_0-\bs{b}'_0).
\label{smallvecb0}
\end{align}

These differential equations imply that $\bs{a}'_0-\bs{b}'_0$ is
$T$-independent at the leading order.
Using \eqref{NN} we find
\begin{align}
&\bs{a}'_0-\bs{b}'_0\sim\alpha_1\alpha_2\frac{2}{1-N^{3,3}}\bs{N}^3
=-\alpha_1\alpha_2\bs{B},
\label{veca0minusb0}\\
&\bs{a}'_0+\bs{b}'_0\sim 0,
\end{align}
in the exact limit of $T\to+0$.
Plugging \eqref{veca0minusb0} back to \eqref{smallveca0} and
\eqref{smallvecb0}, we find the expression for $\bs{a}'_0$ and
$\bs{b}'_0$:
\begin{align}
&\bs{a}'_0
\sim-\frac{\alpha_1\alpha_2}{2}\bs{B}
-\frac{g^2\alpha_1\alpha_2}{4\alpha_3\log(T/|\alpha_3|)}C\bs{B},
\label{a0loopT}\\
&\bs{b}'_0
\sim\frac{\alpha_1\alpha_2}{2}\bs{B}
-\frac{g^2\alpha_1\alpha_2}{4\alpha_3\log(T/|\alpha_3|)}C\bs{B},
\label{b0loopT}
\end{align}
which implies
\begin{align}
\bs{C}\sim-\frac{g^2}{2\log(T/|\alpha_3|)}C\bs{B}.
\label{asympC}
\end{align}
The final result shows \eqref{CC/c}.

\subsection{Evaluation of $g$}
In order to obtain explicit formulas corresponding to \eqref{other2} and
\eqref{other3} for the one-loop diagram in LCSFT, we have to evaluate
the constant $g$, which appeared in Appendix \ref{sec:loop diagram}.
Here, we determine it by computing a one-loop diagram with two
gravitons inserted in two ways, using respectively bosonic LCSFT and
the $\alpha=p^+$ HIKKO string field theory \cite{alpha=p+}, and
comparing their results.
Note that in \cite{KMT} we applied the same method for a one-loop
diagram with two tachyons inserted to determine
$K_T=\mu^4(4\pi)^{-12}|\sqrt{-c}\det'|^{-24}$  or \eqref{cdet}.

Let us consider, in bosonic LCSFT, a contraction of two graviton states
$\langle\zeta_r,-k_r^i|=
\zeta^{ij}_r\langle-k_r^i|a_1^{(r)i}\bar{a}_1^{(r)j},~(r=3,6)$
with
$k_r^i\zeta^{ij}_r=0,
\zeta^{ij}_r=\zeta^{ji}_r,\zeta^{ij}_r\delta_{ij}=0$
\sloppy
and $|B(3,6)\rangle$, which is given by (34) in \cite{KMT}.
It is evaluated as
\begin{align}
\langle\zeta_3,-k_3^i|\langle\zeta_6,-k_6^i|B(3,6)\rangle
\sim\zeta_3^{ij}\zeta_6^{kl}
\left[\delta^{il}\delta^{jk}\left(-\frac{1}{2c}
\bigl(({\boldsymbol C})_1\bigr)^2\right)^2
+\delta^{jl}\delta^{ik}
\right]K_T(2\pi)^{24}\delta^{24}(k_3+k_6),
\label{eq:2gravB}
\end{align}
for $T\to +0$, where $K_T$ appears similarly to the computation of the
one-loop diagram with two tachyons inserted \cite{KMT}.
{}From \eqref{asympC}, we have
\begin{align}
\label{eq:c12g}
-\frac{1}{2c}
\bigl(({\boldsymbol C})_1\bigr)^2
\sim\frac{g^2}{2\pi^2|\alpha_1\alpha_2/\alpha_3|}
\frac{\sin^2(\pi\alpha_1/|\alpha_3|)}{-\log(T/|\alpha_3|)}\,,
\end{align}
which implies that we can determine $g$ from the evaluation of
\eqref{eq:2gravB}.

Including the light-cone directions and the level matching projection,
the total amplitude for the one-loop diagram with two gravitons is
computed as
\begin{align}
T_{36}&=
\langle\zeta_3,-k_3|\langle\zeta_6,-k_6|
\langle R^{{\rm LC}}(2,5)|\langle R^{{\rm LC}}(1,4)|
\Delta_1\Delta_2
|V^{{\rm LC}}(1,2,3)\rangle|V^{{\rm LC}}(4,5,6)\rangle\nonumber\\
&=\int_0^\infty dT\int d\alpha_1
\oint\frac{d\theta_1}{2\pi}\oint\frac{d\theta_2}{2\pi}
(2\pi)^2\delta(k_3^-+k_6^-)\delta(k_3^++k_6^+)
\frac{e^{2Tk_3^-}}{4\pi\alpha_1\alpha_2}
\nonumber\\
&\qquad\qquad\qquad\qquad\qquad\qquad\times 
\langle\zeta_3,-k_3^i|\langle\zeta_6,-k_6^i
|B_{\theta_1,\theta_2}(3,6)\rangle\,,
\label{eq:T36LCSFT}
\end{align}
where $\Delta_r$ is the propagator combined with the level matching
projection:
\begin{align}
\Delta_r=\frac{1}{-2p^+_rp^-_r+L_0^{(r)}+\bar L_0^{(r)}}{\cal P}_r
=\int_0^\infty\frac{dT_r}{\alpha_r}\oint\frac{d\theta_r}{2\pi}
e^{-\frac{T_r}{\alpha_r}(-2p^+_rp^-_r+L_0^{(r)}+\bar L_0^{(r)})}
e^{i\theta_r(L_0^{(r)}-\bar L_0^{(r)})},
\end{align}
and $|B_{\theta_1,\theta_2}(3,6)\rangle$ is the effective interaction
vertex rotated from the original one $|B(3,6)\rangle$ with
$|B_{\theta_1=0,\theta_2=0}(3,6)\rangle=|B(3,6)\rangle$.

The above on-shell amplitude $T_{36}$ can be also obtained in the
framework of the $\alpha=p^+$ HIKKO string field theory:
\begin{align}
T_{36}&=
\langle\zeta_3,-k_3|\langle\zeta_6,-k_6|
\langle R^{\alpha=p^+}(2,5)|\langle R^{\alpha=p^+}(1,4)|
\frac{b_0^{(1)}\bar b_0^{(1)}}
{L_0^{{\rm tot}(1)}+\bar L_0^{{\rm tot}(1)}}
{\cal P}_1^{\rm tot}
\frac{b_0^{(2)}\bar b_0^{(2)}}
{L_0^{{\rm tot}(2)}+\bar L_0^{{\rm tot}(2)}}
{\cal P}_2^{\rm tot}\nonumber\\
&\qquad\qquad\qquad\qquad\qquad\qquad\times
|V^{\alpha=p^+}(1,2,3)\rangle|V^{\alpha=p^+}(4,5,6)\rangle\nonumber\\
&=\int_0^{\infty}dT_1\int_0^{\infty}dT_2
\oint\frac{d\theta_1}{2\pi}\oint\frac{d\theta_2}{2\pi}
\langle\zeta_3,-k_3|\langle\zeta_6,-k_6|
\langle R^{\alpha=p^+}(2,5)|\langle R^{\alpha=p^+}(1,4)|
\frac{1}{\alpha_1\alpha_2}\nonumber\\
&\qquad\times 
b_0^{(1)}\bar b_0^{(1)}b_0^{(2)}\bar b_0^{(2)}
e^{-\frac{T_1}{\alpha_1}(L_0^{{\rm tot}(1)}+\bar L_0^{{\rm tot}(1)})
-\frac{T_2}{\alpha_2}(L_0^{{\rm tot}(2)}+\bar L_0^{{\rm tot}(2)})
+i\theta_1(L_0^{{\rm tot}(1)}-\bar L_0^{{\rm tot}(1)})
+i\theta_2(L_0^{{\rm tot}(2)}-\bar L_0^{{\rm tot}(2)})}
\nonumber\\
&\qquad\qquad\qquad\qquad\qquad\qquad\times
|V^{\alpha=p^+}(1,2,3)\rangle|V^{\alpha=p^+}(4,5,6)\rangle,
\end{align}
which includes the ghost part in addition to the light-cone directions. 
We calculate it using the CFT correlator on the torus $u$-plane
($u\sim u+1\sim u+\tau$):
\begin{align}
T_{36}&=\int_0^{\infty}dT_1\int_0^{\infty}dT_2
\oint\frac{d\theta_1}{2\pi}\oint\frac{d\theta_2}{2\pi}
\bigl\langle
(\alpha_1\alpha_2)^{-1}
b^{(1)}\bar b^{(1)}b^{(2)}\bar b^{(2)}\,
V_{\zeta_6,k_6}(U_6,\bar U_6)
V_{\zeta_3,k_3}(U_3,\bar U_3)
\bigr\rangle_{\tau}\,,
\label{eq:T36corr}
\end{align}
where $b^{(i)}$ is given by a contour integral on $C_i$ in the $u$-plane,
which is denoted in Fig.~3 \cite{KMT} for a pure imaginary $\tau$, and
$V_{\zeta,k}(u,\bar u)$ is the graviton vertex:
\begin{align}
b^{(i)}=\int_{C_i}\frac{du}{2\pi i}\alpha_i\frac{du}{d\rho}b(u),\quad
V_{\zeta,k}(u,\bar u)
=\frac{1}{4}\zeta^{ij}c(u)\bar c(\bar u)\!:\!
i\partial X^i(u)i\bar\partial\bar X^j(\bar u)
e^{ik_{\mu}X^{\mu}(u,\bar u)}\!\!:\,.
\end{align}
Note that here we have used the on-shell condition for the graviton
vertices: $k_3^{\mu}k_{3\mu}=k_6^{\mu}k_{6\mu}=0$.
The light-cone diagram ($\rho$-plane) can be obtained from 
the torus $u$-plane by the generalized Mandelstam map:
\begin{align}
\label{eq:mandel_gen1}
\rho(u)=|\alpha_3|\left(
\log\frac{\vartheta_1(u-U_6|\tau)}{\vartheta_1(u-U_3|\tau)}
+2\pi i\frac{{\rm Im}(U_3-U_6)}{{\rm Im}\tau}u\right),
\end{align}
with $|\alpha_3|=k_3^+$ and $U_3+U_6=0$.
It is related to the parameters on the light-cone diagram as
\begin{align}
&\rho(u+1)-\rho(u)=-2\pi i\alpha_1,
\label{eq:cond_cyc1}
\\
&\rho(u+\tau)-\rho(u)=T_2-T_1-i(\alpha_2\theta_2-\alpha_1\theta_1),
\label{eq:cond_cyc2}
\\
&\rho(u_{-})-\rho(u_{+})=T_2-i\alpha_2\theta_2,
\quad\frac{d\rho(u_{\pm})}{du}=0\,.
\label{eq:cond_intT}
\end{align}
{}From the explicit computation with the $\alpha=p^+$ prescription,
which provides $\delta({\rm Re}(\rho(u+\tau)-\rho(u)))=\delta(T_2-T_1)$
in the integrand, (\ref{eq:T36corr}) can be rewritten as
\begin{align}
T_{36}&=\int_0^{\infty}dT\int d\alpha_1
\oint\frac{d\theta_1}{2\pi}
\oint\frac{d\theta_2}{2\pi}
(2\pi)^{26}\delta^{26}(k_6+k_3)
\frac{\alpha_1\alpha_2}{2^{38}\pi^{25}\alpha_3^4}
\nonumber\\
&\qquad\times
\Bigl|g_1'(u_+-U_6|\tau)-g_1'(u_+-U_3|\tau)\Bigr|^{-2}
\Bigl|({\rm Im}\tau)^{\frac{1}{4}}\eta(\tau)\Bigr|^{-48}
\nonumber\\
&\qquad\times 
\zeta_3^{ij}\zeta_6^{kl}
\biggl[
\delta^{il}\delta^{jk}\biggl(\frac{\pi}{{\rm Im}\tau}\biggr)^2
+\delta^{jl}\delta^{ik}
\biggl|\frac{\pi}{{\rm Im}\tau}+g_1'(U_6-U_3|\tau)\biggr|^2
\biggr]\,,
\label{eq:T36f}
\end{align}
with $g_1'(\nu|\tau)=\partial^2_{\nu}\log\vartheta_1(\nu|\tau)$.
The factors in the 
second and the third line are functions of two complex parameters 
$\tau,U_3-U_6$, which are related to 4 real parameters
$T(=T_1=T_2),\alpha_1,\theta_1,\theta_2$ by (\ref{eq:cond_cyc1}),
(\ref{eq:cond_cyc2}) and (\ref{eq:cond_intT}).

Similarly, for the one-loop amplitude with two tachyons inserted we
find\footnote{
Because we have included the level matching projection which was omitted
in \cite{KMT}, we can reproduce a modular invariant measure by computing
the Jacobian using \eqref{eq:cond_cyc1}, \eqref{eq:cond_cyc2} and
\eqref{eq:cond_intT}:
\begin{align}
\int_0^{\infty}dT\int \!d\alpha_1\!
\oint\frac{d\theta_1}{2\pi}\!
\oint\frac{d\theta_2}{2\pi}
\frac{\alpha_1\alpha_2}{\alpha_3^4}(\cdots)
=\frac{1}{2\pi}\!\int\! d^2\tau dx dy
\left|g_1'(u_+-U_6|\tau)-g_1'(u_+-U_3|\tau)\right|^2(\cdots).
\end{align}
Here $x,y$ are real parameters defined by $U_3-U_6=x+y\tau$.
However, we have only to use the expression of the integrand for
$\theta_1=0,\theta_2=0$, namely without projection,
in the evaluation of $K_T~(T\to +0)$ in \cite{KMT} and $g$ here.
}
\begin{align}
S_{36}&=\langle -k_3|\langle -k_6|
\langle R^{\alpha=p^+}(2,5)|\langle R^{\alpha=p^+}(1,4)|
\frac{b_0^{(1)}\bar b_0^{(1)}}{L_0^{(1)}+\bar L_0^{(1)}}
{\cal P}_1^{\rm tot}
\frac{b_0^{(2)}\bar b_0^{(2)}}{L_0^{(2)}+\bar L_0^{(2)}}
{\cal P}_2^{\rm tot}\nonumber\\
&\qquad\times
|V^{\alpha=p^+}(1,2,3)\rangle|V^{\alpha=p^+}(4,5,6)\rangle\nonumber\\
&=\int_0^{\infty}dT\int d\alpha_1
\oint\frac{d\theta_1}{2\pi}\oint\frac{d\theta_2}{2\pi}
(2\pi)^{26}\delta^{26}(k_6+k_3)
\frac{\alpha_1\alpha_2}{2^{38}\pi^{25}\alpha_3^4}
\label{eq:S36}\\
&\qquad\times
\Bigl|g_1'(u_+-U_6|\tau)-g_1'(u_+-U_3|\tau)\Bigr|^{-2}
\Bigl|({\rm Im}\tau)^{\frac{1}{4}}\eta(\tau)\Bigr|^{-48}
\left|e^{\pi\frac{({\rm Im}(U_3-U_6))^2}{{\rm Im}\tau}}\frac{
\partial_{\nu}\vartheta_1(\nu|\tau)|_{\nu=0}}
{\vartheta_1(U_6-U_3|\tau)}\right|^4.\nonumber
\end{align}
For $\theta_1=\theta_2=0,T\to +0$, the extra factor of the integrand in
\eqref{eq:T36f} compared to that in \eqref{eq:S36} can be evaluated as 
\begin{align}
&\zeta_3^{ij}\zeta_6^{kl}
\biggl[
\delta^{il}\delta^{jk}\biggl(\frac{\pi}{{\rm Im}\tau}\biggr)^2
+\delta^{jl}\delta^{ik}
\biggl|\frac{\pi}{{\rm Im}\tau}+g_1'(U_6-U_3|\tau)\biggr|^2
\biggr]\times
\left|e^{\pi\frac{({\rm Im}(U_3-U_6))^2}{{\rm Im}\tau}}
\frac{\partial_{\nu}\vartheta_1(\nu|\tau)|_{\nu=0}}
{\vartheta_1(U_6-U_3|\tau)}\right|^{-4}\nonumber\\
&\qquad\qquad
\sim\zeta_3^{ij}\zeta_6^{kl}
\biggl[\delta^{il}\delta^{jk}
\biggl(\frac{\sin^2(\pi\alpha_1/|\alpha_3|)}
{-\log(T/|\alpha_3|)}\biggr)^2
+\delta^{jl}\delta^{ik}
\biggr].
\end{align}
Comparing this factor with \eqref{eq:c12g} and \eqref{eq:2gravB}, we
finally obtain
\begin{align}
\label{eq:gana}
g=\sqrt{2}\pi|\alpha_1\alpha_2/\alpha_3|^{1/2}.
\end{align}

\section{Formulas for the gamma matrices}
Here we would like to prove the formulas \eqref{Gvalue} and
\eqref{Gdiff} first.
The point is to find a recursion relation.
Multiplying \eqref{sandwich} by $\hat\gamma^{d_1\cdots d_r}$ and using
the gamma matrix product formula (See e.g. \cite{Fujii})
\begin{align}
\hat\gamma^{a_1\cdots a_p}\hat\gamma^{b_1\cdots b_q}
=\sum_{k=0}^{\min(p,q)}
\frac{(-1)^{pk-\frac{1}{2}k(k+1)}p!q!}{(p-k)!(q-k)!k!}
\delta^{[a_1}_{[b_1}\cdots\delta^{a_k}_{b_k}
\hat\gamma^{a_{k+1}\cdots a_p]}{}_{b_{k+1}\cdots b_q]},
\end{align}
where $[\cdots]$ denotes the anti-symmetrization of the indices,
we find the recursive relations for small $r$:
\begin{align}
&G^{c_1\cdots c_m,d}_p
=d_{p,m}\hat\gamma^{c_1\cdots c_m}\hat\gamma^d
-d_{p-1,m}\hat\gamma^d\hat\gamma^{c_1\cdots c_m}
+G^{c_1\cdots c_m,d}_{p-2},\\
&G^{c_1\cdots c_m,d_1d_2}_p-\tilde G^{d_1d_2,c_1\cdots c_m}_p
=d_{p,m}[\hat\gamma^{c_1\cdots c_m},\hat\gamma^{d_1d_2}]
+G^{c_1\cdots c_m,d_1d_2}_{p-2}-\tilde G^{d_1d_2,c_1\cdots c_m}_{p-2}.
\end{align}
Using these formula recursively, we can prove \eqref{Gvalue} and
\eqref{Gdiff} without difficulty.

For the explicit computation of $G^{c_1\cdots c_m,d}_p$ and
$G^{c_1\cdots c_m,d_1d_2}_p-\tilde G^{d_1d_2,c_1\cdots c_m}_p$, we need
several summation formulas of $d_{p,m}$.
For this purpose first we note that $d_{p,m}$ has the residue formula:
\begin{align}
d_{p,m}
=(-1)^{pm}\oint_{z=0}{dz\over 2\pi i}z^{-1-p}(1+z)^{8-m}(1-z)^m
=\sum_{s=0}^{\min(p,m)}
{(-1)^s(8-m)!m!\over(p-s)!(8-m-p+s)!s!(m-s)!}.
\end{align}
Using this expression we find we can show the following formulas.
\begin{align}
&\sum_{q=0}^4d_{2q,m}=128(\delta_{m,0}+\delta_{m,8}),
\label{sumd1}\\
&\sum_{q=0}^3d_{2q+1,m}=128(\delta_{m,0}-\delta_{m,8}),
\label{sumd2}\\
&\sum_{q=0}^3\sum_{q_0=0}^qd_{2q_0+1,m}
=320(\delta_{m,0}-\delta_{m,8})-32(\delta_{m,1}-\delta_{m,7}),
\label{sumd3}\\
&\sum_{q=0}^3\sum_{q_0=0}^qd_{2q_0,m}
=256(\delta_{m,0}+\delta_{m,8})+32(\delta_{m,1}+\delta_{m,7}),
\label{sumd4}\\
&\sum_{q=0}^2\sum_{q_0=0}^qd_{2q_0+1,m}
=192(\delta_{m,0}-\delta_{m,8})-32(\delta_{m,1}-\delta_{m,7}).
\label{sumd5}
\end{align}

\end{document}